\documentstyle[aps,multicol,prb,epsf]{revtex}

\newcommand\r{{\bf r}}
\newcommand\n{{\bf n}}

\begin{document}
\ifpreprintsty \baselineskip=18pt \else\fi \bibliographystyle{simpl1}
\date{\today}

\title{The Ehrenfest Oscillations in The Level Statistics of Chaotic 
Quantum Dots}

\author{Chushun Tian$^1$ and Anatoly I. Larkin$^{1,2}$}
\address{$^1$ William I. Fine Theoretical Physics Institute, University of Minnesota, 
Minneapolis, MN 55455\\
$^2$L.D.Landau Institute for Theoretical Physics, Moscow, 117940, Russia}

\maketitle
\begin{abstract}

We study a crossover from classical to quantum picture in the electron energy statistics in 
a system with broken time-reversal symmetry. 
The perturbative and nonperturbative parts of the two level 
correlation function, $R(\omega)$ are analyzed.  We find that in the intermediate region, 
$\Delta\ll\omega\sim t_E^{-1}\ll t_{erg}^{-1}$, where $t_E$ and $t_{erg}$ are the Ehrenfest and ergodic times, respectively, 
$R(\omega)$ consists of a series of oscillations with the periods depending 
on $t_E$, deviating from the universal Wigner-Dyson statistics. 
These Ehrenfest oscillations have the period 
dependence as $t_E^{-1}$ in the perturbative part. [For systems with time-reversal symmetry, 
this oscillation in the perturbative part of $R(\omega)$ was 
studied in an earlier work (I. L. Aleiner and A. I. Larkin, Phys. Rev. E {\bf 55}, 
R1243 (1997))]. In the 
nonperturbative part they have the period dependence as $\left(\Delta^{-1}+\alpha t_E\right)^{-1}$ with $\alpha$
a universal numerical factor. The amplitude of the leading order Ehrenfest oscillation in the 
nonperturbative part is larger than that of the perturbative part.
\end{abstract}
\draft
\vspace*{-5mm}

\ifpreprintsty\else \begin{multicols}{2} \fi

\section{Introduction}
\label{sec:1}

In the last decade, there has been increasing interest in investigating 
interference phenomena of electronic motion in ballistic quantum dots 
\cite{Bar98}. In semiconductor quantum dots the confining potential has the length scale much larger than the Fermi
wavelength, and therefore one may expect that electron motion in such systems 
may be described by methods of classical trajectories. Typically, the classical 
electron motion inside dots is chaotic. An important question is, what happens to the 
energy spectrum. It turns out that the energy spectrum of a single electron in 
chaotic quantum dots is highly  sensitive to 
the parameters of the system (e.g., Fermi 
momentum, the strength of the applied magnetic field, the impurity 
configuration, the shape of the boundary of ballistic quantum dots, etc.). 
Substantial progress has been made by studying the 
statistics of energy levels. The level statistics has played an important role in the theories of atomic 
nuclei\cite{Wigner,Porter}, disordered metals\cite{GE65,Efetov83,AS86}, and 
quantum chaos\cite{Gut90}. A particularly interesting quantity in the level 
statistics is the so-called two level correlation function, $R(\omega)$:
\begin{equation}
R(\omega)=\Delta^2\langle
{\rm Tr}(\epsilon+\frac{\omega}{2}-{\hat H})
{\rm Tr}(\epsilon-\frac{\omega}{2}-{\hat H})\rangle
\label{Def_R}
\end{equation}
where $\Delta$ is the mean level spacing and ${\hat H}$ is the 
Hamiltonian. 
$\langle\cdots\rangle$ denotes the average over a wide energy band (for 
individual 
quantum dots) or the impurity configuration (for quantum disorders).

In his pioneering work, Dyson divided the Hamiltonian matrices into 
three classes: Gaussian orthogonal ensemble (GOE), Gaussian unitary
ensemble (GUE) and Gaussian sympletic ensemble (GSE) and used the random 
matrix theory (RMT) to find that 
$R(\omega)$ 
follows different universal behavior for these ensembles \cite{Meh90,Dyson60}. Gor'kov and Eliashberg were 
the first to use Wigner-Dyson ensemble to 
study transport properties in small metallic grains\cite{GE65}. For 
disordered metals, the 
behavior of $R(\omega)$ is now well understood. For grains in the 
presence of a lot of quantum (Born) impurities, theoretical 
justification of the 
universal RMT result came up with the supersymmetric field theory
due to Efetov \cite{Efetov83}, assuming that $\omega$ is much smaller 
than the Thouless 
energy $E_{Th}=\hbar D/L^2$ with $D$ being the diffusion constant and 
$L$ the size of the grain. The behavior of $R(\omega)$ beyond $E_{Th}$ is 
nonuniversal \cite{AS86}. That is, the level statistics at $\omega\gtrsim E_{Th}$ 
is system-dependent. In Ref.~\onlinecite{KM94}, it was found 
that there is a small constant correction to $R(\omega)$ at 
$\omega\sim \Delta$ due to 
the nonuniversal behavior of electron motion. Recently, the weak localization correction to $R(\omega)$ beyond the 
Thouless energy $E_{Th}$ 
was 
studied in Ref.~\onlinecite{SLA98} with the help of replica technique 
\cite{ELK80}. 

Bohigas, Giagonni, and Schmit (BGS)
proposed a 
conjecture that the fluctuations of levels of quantum chaotic systems may follow the universal 
RMT results \cite{BGS84}. One routine method to study 
the behavior of $R(\omega)$ in 
ballistic quantum dots is to employ the 
Gutzwiller 
formula \cite{Gut71}. A cornerstone was put by Berry \cite{Ber85}. By 
taking 
into account the diagonal contributions, he was able to use the 
Hannay-Ozorio de Almeida (HOA) sum rule \cite {HOA84} to reproduce the 
leading perturbative term of the universal RMT results. Proceeding 
along this line, an important progress was made in 
Ref.~\onlinecite{BK96}, where the off-diagonal contributions 
were studied and the oscillation similar to the universal RMT results 
was thereby found.

In this paper, we will study the ballistic quantum dot with weak diffractions, 
where the classical electronic motion is chaotic (chaotic quantum dots). In such case, there is a new scale $t_E$, so-called 
Ehrenfest 
time, which has logarithmic dependence on $\hbar$:
\begin{equation}
t_E=\frac{1}{\lambda}|\ln \left(\frac{L}{\lambda_F}\right)|\sim 
\frac{1}{\lambda} |\ln \hbar|.
\label{Ehrenfest time}
\end{equation}
Here $\lambda$ is the Lyapunov exponent, $L$ is some macroscopic size 
and $\lambda_F$ is the Fermi wavelength. Physically, it is the time 
scale for an 
initial Gaussian wave packet (with the typical spatial spreading 
$\sim\lambda_F$) to develop to 
some macroscopic size. At this scale, the electron motion shows a crossover 
from classical to quantum picture. Such a crossover exists in other phenomena also. Larkin and Ovchinikov \cite{LYu68} found 
that it is 
possible to use the method of classical trajectories in the theory of 
superconductor 
for $t<t_E$ only. Besides, they were able to estimate $t_E$ for 
Lorentz gases of semiclassical impurities. 
The logarithmic dependence on $\hbar$ of $t_E$ was also found by Berman and Zaslavsky in the 
studies of classical chaos of the kicked rotator model\cite{Zasl60}. In the 
past two decades, it has been found that the Ehrenfest time plays an 
important role in condensed matter. Wilkinson addressed the 
importance of such a scale 
in the semiclassical studies of sum rules over generic matrix elements 
\cite{Wil87}. Argaman studied the 
conductance in the scale $\omega\ll t_E^{-1}$~\cite{Arg95}.  In 
Ref.~\onlinecite{AL96}, 
Aleiner and Larkin not only gave the estimation for $t_E$, but also 
studied 
the weak localization correction to the conductivity at the crossover scale $\omega\sim 
t_E^{-1}$. 
They showed that diffusons do not couple at time $t<t_E$ while at 
time $t\gtrsim t_E$, the coupling reaches a universal value. As a 
result, the 
amplitude of the coupling between diffusons (Hikami box) 
\cite{Hikami80} is an 
oscillating function of $\omega t_E$. Later it was found that 
such amplitude has important applications in studying the crossover 
from classical and quantum picture of shot noise 
\cite{AAL00,TTSB03} 
and the density of states (DOS) in Andreev billiard 
\cite{VL02,JSB03}.

For chaotic quantum dots with one macroscopic scale $L$,  the ergodic 
time $t_{erg}\sim L/v_F$, which is of the same order of $\lambda^{-1}$. 
In the semiclassics, $t_E$ opens a new region: $\Delta\ll 
t_E^{-1}\sim \omega\ll t_{erg}^{-1}$. 
The manifestation of $t_E$ in the level statistics of chaotic quantum dots was first studied in 
Ref.~\onlinecite{AL97}. According to the RMT, asymptotically, for large $\omega$ the two level correlation 
function consists of two parts: One part is perturbative in $\frac{\Delta}{\omega}$. The other is nonperturbative, which oscillates with the period $\Delta$. In Ref.~\onlinecite{AL97}, 
it was found that there is a small oscillation with the period proportional to 
$t_E^{-1}$ correcting the perturbative part of the universal 
$R(\omega)$ for GOE \cite{AL97'}:
\begin{equation}
\Delta R^{\rm o}_p(\omega)=
\frac{\Delta^3}{2\pi^3}{\rm Re}\frac{\partial^2}{\partial(-i\omega)^2}
\frac{\Gamma_2^2(\omega)}{-i\omega},
\label{Intro_1}
\end{equation}
where $\Gamma_2(\omega)$ is 
\begin{equation}
\Gamma_2(\omega)=\exp \left(i\omega t_E-\frac{\omega^2\lambda_2 
t_E}{\lambda^2}\right).
\label{Gamma2}
\end{equation} 
Here $\lambda$ is the Lyapunov exponent and $\lambda_2$ characterizes 
the fluctuations of $\lambda$ [see Eqs.~(\ref{aaa}) and (\ref{aaa'}) for the 
general definitions]. Typically, $\lambda$ and $\lambda_2$ are 
of the same order.

In this paper, we study the two level correlation function in the system 
with broken time-reversal symmetry. We recall that in this case, the RMT predicts the famous result that the 
two level correlation function is described by the elegant formula:
\begin{equation}
R^{\rm u}(\omega)=
1-\frac{\Delta^2}{2\pi^2\omega^2}\left(1-\cos \frac{2\pi\omega}{\Delta}\right).
\label{Intro_3}
\end{equation}
Asymptotically, for large $\omega$, the two level correlation function 
is exactly truncated at the $(\Delta/\omega)^2$ term \cite{Efetov83,Meh90}. 
An important question is whether it is the property of generic systems with 
broken time-reversal symmetry or just comes out with the RMT for GUE. 
In this paper, we will show that Eq.~(\ref{Intro_3}) just comes out with RMT, or describes 
the universal limit of the two level correlation function in quantum disorders. 
For ballistic quantum dots, surprisingly, we find that in the crossover region 
$t_E^{-1}\sim \omega\ll t_{erg}^{-1}$, 
the behavior of the two level correlation function deviates from the universal Wigner-Dyson 
statistics given by Eq.~(\ref{Intro_3}). 
More exactly, such deviations are formulated as
\begin{equation}
\Delta R^{\rm u}(\omega)=\Delta R^{\rm u}_p(\omega)+
\Delta R^{\rm u}_{np}(\omega),
\label{Intro_2}
\end{equation}
where 
\begin{equation}
\Delta R^{\rm u}_p(\omega)=
-\frac{\Delta^4}{8\pi^4} {\rm Re} 
\frac{\partial^2}{\partial(-i\omega)^2} 
\left[\frac{\Gamma_2^4(\omega)}{(-i\omega)^2}
-\frac{\Gamma_3^2(\omega)}{(-i\omega)^2}\right],
\label{RuniEO}
\end{equation}
and 
\begin{equation}
\Gamma_3(\omega)=
\exp \left(
\frac{3}{2}i\omega t_E-\frac{9}{4}\frac{\omega^2\lambda_2 
t_E}{\lambda^2}
\right).
\label{Gamma33}
\end{equation} 
\begin{equation}
\Delta R^{\rm u}_{np}(\omega)=2\frac{\Delta^3}{\pi^3\omega^3}
\sin \frac{2\pi \omega}{\Delta}
{\rm Re}
\left[
\Gamma_2^2(\omega)-\omega
\frac{\partial}{\partial \omega}\Gamma_2^2(\frac{\omega}{2})-1
\right].\label{Intro_2_np}
\end{equation}

Compared to Eq.~(\ref{Intro_1}), the deviations in systems with broken time-reversal symmetry are more interesting. 
First, we point out that the main feature of these deviations 
is the appearance of 
two types of oscillations in addition to the oscillation described by the Wigner-Dyson statistics. 
One type correcting the perturbative terms has the period 
$\sim t_E^{-1}$, which has the similar origin as Eq.~(\ref{Intro_1}). In fact, 
as we will show in Sec.~\ref{sec3}, Eq.~(\ref{Intro_1}) comes from the two loop 
approximation. In the broken time-reversal system, the two loop approximation 
vanishes. Instead, Eq.~(\ref{RuniEO}) comes 
from the three loop approximation. They are the weak localization corrections to 
leading universal $R(\omega)$ in the crossover region. 
Eq.~(\ref{Intro_2_np}) belongs to the other type 
of oscillation, which corrects the 
nonperturbative part of the universal $R(\omega)$ described by the universal Wigner-Dyson statistics 
and was not found in the previous work \cite{AL97}. It has the period 
$\left(\Delta^{-1} +\alpha t_E \right)^{-1}$. Both of them have $t_E$-dependent periods. In this way, we term them 
the Ehrenfest oscillations. 
In the broken time-reversal system, the oscillation 
with the period $\left(\Delta^{-1} +\alpha t_E \right)^{-1}$ is stronger than 
the oscillation with period $\sim t_E^{-1}$. For the time-reversal system, 
the Ehrenfest oscillation with the period $\left(\Delta^{-1} +\alpha t_E \right)^{-1}$ 
does exist, but the amplitude is smaller than that described by the 
Wigner-Dyson statistics, which is even small.

In Ref.~\onlinecite{ASAA96}, it is proposed that the amplitudes of both perturbative 
and nonperturbative parts of $R(\omega)$ are related to each other through the classical spectrum 
determinant, which depends on the eigenvalues of the Perron-Frobenius operator. 
In the universal limit $\omega\ll t_{erg}^{-1}$, it coincides with the result of 
RMT. Indeed, for the broken time-reversal system, we see from Eq.~(\ref{Intro_3}) that 
the amplitudes of both parts are in the same order. However, such a conjecture 
does not exist once the quantum corrections are taken into account. In fact, 
the leading quantum correction in the perturbative part, i.e., Eq.~(\ref{RuniEO}) comes from the three loop approximation, while 
the leading quantum correction in the nonperturbative part, i.e., Eq.~(\ref{Intro_2_np}) 
comes from the two-loop approximation. As a result, the leading order quantum corrections 
lead to the Ehrenfest oscillations with different power in $\Delta/\omega$.

It is worth pointing out that the methods of calculating coupled multi-diffusons 
(see Sec.~\ref{sec4}) involved in the three loop 
approximation are general. The idea may be applied to find the weak localization 
correction to other physical quantities (e.g. conductivity) in the ballistic quantum 
dots.

Technically, it is convenient to use the ballistic nonlinear 
supermatrix $\sigma$ model (BNS$\sigma$M) \cite{ASAA96,MKh95,KE02}. 
We will introduce this model in the next section. The
quantum transport time, essential to the regularizer will be estimated. 
The supersymmetric action has two saddle points: $\Lambda$ and 
$-k\Lambda$ 
\cite{AA95}. The perturbative expansion around $\Lambda$ gives 
the weak localization corrections to the leading perturbative term in the Wigner-Dyson 
statistics, while the perturbative expansion around $-k\Lambda$ gives 
the quantum corrections to the leading Wigner-Dyson oscillation. In 
Sec.~\ref{sec3}, we consider the perturbative $R(\omega)$ for systems with time-reversal symmetry.  
The perturbation theory near $\Lambda$ will be developed. 
In Sec.~\ref{sec4}, we consider systems with broken time-reversal symmetry. The perturbation theory 
near $\Lambda$ will be employed to study the weak localization corrections in $R_p(\omega)$. 
The products of the perturbative expansion involve the product of multi-diffusons. We will 
calculate such coupled diffusons. Immediately, the leading order quantum corrections give
the Ehrenfest oscillations in $R_p^{\rm u}(\omega)$. In Sec.~\ref{sec5}, with the help of the global transformation, we will perform the 
perturbation expansion near $-k\Lambda$. Consequently, we find the Ehrenfest oscillations 
in the nonperturbative part of $R(\omega)$ but with different dependence of the period on $t_E$. 
The obtained results are summarized in Sec.~\ref{sec7}. Some of the calculations of 
multi-diffusons are included in Appendix~A.

\section{Ballistic Nonlinear Supermatrix $\sigma$ Model (BNS$\sigma$M)}
\label{sec:2}

Diffusive nonlinear supermatrix $\sigma$ model has become a powerful 
tool in the studies of $R(\omega)$ in quantum disorders 
\cite{Efetov83,AA95}. A 
natural question is where such a technique can be generalized to 
the 
ballistic case, especially individual quantum dots, whether the average 
over 
impurity configurations become impossible. Towards this direction, the 
first 
suggestion is given in Ref.~\onlinecite{MKh95}, where a ballistic 
action was 
phenomenologically proposed. The formal justification is obtained in 
Ref.~\onlinecite{ASAA96} under some crucial assumptions \cite{AOB}. The complete 
microscopic derivation is 
given for long-ranged disorders \cite{KE02,EL03}. In this 
section, 
we will introduce BNS$\sigma$M. Moreover, the roles played by the regularizer will be 
discussed.

\subsection{Ballistic supersymmetric action}
\label{sec2.1}

It is a standard method to introduce a supermatrix field $Q$. It is 
defined on the superspace $p\otimes g\otimes d$, where $p$ stands for 
the advanced/retarded block, $g$ stands for the fermion/boson block and 
$d$
stands for the time-reversal block. Similarly to the quantum disordered 
case \cite{Efetov83}, 
$R(\omega)$, Eq.~(\ref{Def_R}) can be expressed as the integral over $Q$:
\begin{eqnarray}
R(\omega)&=&{\rm Re} {\tilde R}(\omega), \nonumber\\
{\tilde R}(\omega)&=&\frac{1}{64}\int DQ \left(\int dx_\parallel {\rm STr}
k\Lambda Q(x_\parallel)\right)^2 e^{-S}.
\label{R}
\end{eqnarray}
Here the subscript in $x_\parallel$ means that the integration is 
restricted 
on the energy shell. $\Lambda$ is a supermatrix defined as:
\begin{equation}
\Lambda=\left(\matrix{1&0\cr0&-1}\right)_p\otimes 1_g\otimes 1_d.
\label{Lambda}
\end{equation}
The ballistic supersymmetric action $S$ is:
\begin{equation}
S=\frac{\pi\nu}{2}\int d1
{\rm STr}\left[\frac{i\omega^+}{2}\Lambda 
Q-T^{-1}\Lambda
{\hat {\cal L}}T+\frac{1}{\tau_q}\left(
\frac{\partial Q}{\partial\phi }\right)^2\right],
\label{SUSY action}
\end{equation}
where $\omega^+=\omega+i0^+$ with $0^+$ being an arbitrarily small positive number. $1=(\r,\n), d1=d\r d\n /2\pi$. 
${\hat {\cal L}}={\bf v} \cdot \nabla-\nabla 
U(\r)\cdot\frac{\partial}{\partial {\bf p}}$ 
is the Liouville operator with $U(\r)$ being the classical potential. 
The matrix $Q$
is generated by $T$:
\begin{equation}
Q=T^{-1}\Lambda T,
\label{Q}
\end{equation}       
and takes the value on the corresponding symmetry 
space 
${\bf H}={\bf G}/{\bf K}$, where ${\bf G}$ and ${\bf K}$ are groups. 
For GUE, 
${\bf H}={\bf U}(1,1/2)/{\bf U}(1/1)\otimes {\bf 
U}(1/1)$. For GOE, ${\bf H}={\bf U}(2,2/4)/{\bf U}(2/2)\otimes {\bf 
U}(2/2)$. $T(1)$ satisfies:
\begin{equation}
T^\dag(1)LT(1)=L,
\label{Tcon}
\end{equation}
where $L$ is a matrix defined as
\begin{equation}
L=\left(\matrix{1&0\cr0&k\cr}\right)_p\otimes 1_d, \qquad
k=\left(\matrix{1&0\cr0&-1\cr}\right)_g.
\label{Lmat}
\end{equation}

The form of the regularizer is determined by the properties of the Liouville operator 
and diffraction mechanisms for particular systems. In the presence of quantum impurities with small-angle scattering, the 
regularizer has the form presented in Eq.~(\ref{SUSY action}) with 
$1/\tau_q$ 
proportional to the impurity density \cite{AL96,AL97}. In this paper, we will use it to regularize the Liouville operator. According to the geometric 
theory of diffractions \cite {Kel62}, when an electron glances off a hard 
disk with the radius $\rho$, it emits diffraction rays into the shadowed region 
deviating from the incident ray at the angle 
$\theta\sim\left(\frac{\lambda_F}
{\rho}\right)^{2/3}$ \cite{VWR94}. This leads to the estimation: 
$\frac{1}{\tau_q}\sim\frac
{v_F}{L}\left(\frac{\lambda_F}{\rho}\right)^{2/3}$, which coincides 
with the estimation in Ref.~\onlinecite{SU}. For smooth 
potentials, it is estimated that $\tau_q^{-1}\sim v_F\lambda_F/L^2$ 
~\cite{LYu68,AL96}, which is also obtained in Ref.~\onlinecite{VL02} 
by analyzing the spreading of the wave packet. In any cases, 
$\tau_q\sim\hbar^\alpha$. 
It is important that the regularizer is not zero but small. In Ref.~\onlinecite{AL96} (see also Sec.~\ref{sec2.4}), 
it was shown that the 
Ehrenfest time 
$t_E=\frac{1}{\lambda}\ln\lambda\tau_q$. For $\lambda\tau_q\gg 1$, 
$t_E$ 
depends on $\tau_q$ logarithmically. In the semiclassics, $\tau_q$ is large but not infinite. 
Any change of the form of the 
regularizer accounts only for the value of the parameter $\tau_q$ under 
logarithm.

\subsection{The perturbation theory near $\Lambda$}
\label{PerGOE}

To integrate over all the modes in Eqs.~(\ref{R}) and (\ref{SUSY 
action}), we will use the saddle point approximation. 
In quantum disorders, the Gaussian approximation leads to 
Altshuler and Shklovskii's result \cite{AA95}. For $\omega<t_{erg}^{-1}$, only the zero mode is important, which is coordinate 
independent. The coupling between diffusive modes accounts for the higher 
order term 
in the saddle point approximation. Similarly, in chaotic quantum dots, 
the the eigenmodes of the regularized Liouville operator may contribute to $R(\omega)$ independently in the Gaussian approximation. Moreover, 
there is coupling between these eigenmodes also.

The action, Eq.~(\ref{SUSY action}) has two saddle points: 
$\Lambda$ 
and $-k\Lambda$ \cite{AA95}. In Sec.~\ref{sec3} and \ref{sec4}, we will study the 
case with $Q$ close to $\Lambda$. As in the 
disordered case, $R_p(\omega)$ is dominated by this saddle 
point \cite{ASAA96,AA95}. It is convenient to choose the 
parameterization below for $T$:
\begin{equation}
T=1+iP.
\label{parameterization}
\end{equation}
The Jacobian of this transformation is unity. $P$ anticommutes with 
$\Lambda$:
\begin{equation}
\Lambda P+P\Lambda=0,\quad P=\left(
\begin{array}{cc}
 0&B\\
 {\bar B}&0
 \end{array}
 \right).
\label{P}
\end{equation}
According to Eq.~(\ref{Tcon}), $P(1)$ satisfies the following 
condition:
\begin{equation}
P(1)^\ast =-CP({\bar 1})C^T,
\label{condition}
\end{equation}
where ${\bar 1}=(\r, -\n)$ and $C$ is the charge conjugation matrix:
\begin{equation}
C=1_p\otimes\left(
\begin{array} {cc}
 -i\tau_2&0\\
 0&\tau_1
 \end{array}
 \right)_g.
\label{C}
\end{equation}
The Pauli matrices:
 \ifpreprintsty \else\end{multicols} 
                  	\vspace*{-3.5ex} \hspace*{0.491\hsize} 
			\begin{minipage}{0.48\hsize}$\,$\hrule
			 \end{minipage}
\fi
\begin{equation}
\tau_0=\left(
\begin{array}{cc} 
1&0\\ 0&1
\end{array}
\right)_d,\qquad
\tau_1=\left(
\begin{array}{cc} 
0&1\\ 1&0
\end{array}
\right)_d,\qquad
\tau_2=\left(
\begin{array}{cc} 
0&-i\\ i&0
\end{array}
\right)_d,\qquad
\tau_3=\left(
\begin{array}{cc} 
1&0\\ 0&-1
\end{array}
\right)_d,
\label{Pauli matrices}
\end{equation}
 	\ifpreprintsty \else 
                  	\vspace*{-3.5ex} 
			\begin{minipage}{0.48\hsize}$\,$\hrule
			 \end{minipage}\vspace*{1.5ex}
                  \begin{multicols}{2}
                   \fi\noindent
act on the time-reversal block. Under the parameterization 
Eq.~(\ref{parameterization}), we 
have the following expansion:
\begin{eqnarray}
{\rm STr}\left[k\Lambda Q\right]=2\sum_{j=1}^\infty {\rm STr}\left[k 
(-1)^jP^{2j}\right],\label{pre_expan1}\\
{\rm STr}\left[\Lambda Q\right]=2\sum_{j=1}^\infty {\rm STr}
\left[(-1)^jP^{2j}\right],
\label{pre_expan2}
\end{eqnarray}
and
\ifpreprintsty \else\end{multicols} 
                  	\vspace*{-3.5ex} \hspace*{0.491\hsize} 
			\begin{minipage}{0.48\hsize}$\,$\hrule
			 \end{minipage}
\fi
\begin{eqnarray}
S&=&S^0+S^{int.},\label{act_expan}\\
S^0&=&\frac{\pi\nu}{2}\int dx_{\parallel} {\rm 
STr}\left[P(-i\omega^++\Lambda\hat{{\cal L}}_R)P\right],\label{S0}\\
S^{int.}&=&-\frac{\pi\nu}{2}\int dx_{\parallel} \sum_{j=1}^\infty
{\rm STr}\left[(-1)^{j+1}P^{2j+1}(-i\omega^++\Lambda
\hat{{\cal L}}_R)P\right].
\label{intOrt}
\end{eqnarray}
\ifpreprintsty \else 
                  	\vspace*{-3.5ex} 
			\begin{minipage}{0.48\hsize}$\,$\hrule
			\end{minipage}\vspace*{1.5ex}
                  \begin{multicols}{2}
                   \fi\noindent
Here the Perron-Frobenius operator ${\hat {\cal L}}_R$ is defined as \cite{Ruelle86}:
\begin{equation}
{\hat {\cal L}}_R=
{\bf v}\cdot\nabla-\nabla U({\bf r})\cdot\frac{\partial}{\partial {\bf 
p}}-\frac{1}{\tau_q}\frac{\partial^2}{\partial \phi^2}.
\label{FPopera}
\end{equation}
 These expansions are essential to the perturbation calculations.

Keeping the terms up to $P^2$, we reproduce the leading terms of the 
universal 
Wigner-Dyson results for 
$R_p(\omega)$ [see Eqs.~(\ref{zeroRo}) and (\ref{zeroRu})]. In Sec.~\ref{sec3} and \ref{sec4}, we will 
calculate the weak localization correction 
to $R_p(\omega)$. The feature of finite $t_{erg}$ implies the existence of the gap in the spectrum of 
Perron-Frobenius operator, which has been found in many chaotic systems \cite {Gas96,Beck}. 
we will drop out the $P^{2j+1}\Lambda
\hat{{\cal L}}_RP$ term in the effective interaction $S^{int.}$ hereafter since it gives much smaller contribution at $\omega\ll t_{erg}^{-1}$.

\subsection{Quantum disorders}
\label {2.2}

Now we discuss the $\tau_q$ dependence 
of $R(\omega)$. In the presence of many quantum 
impurities 
(i.e., quantum disorders), $\tau_q$ becomes very small ($v_F/L\ll 
1/\tau_q$). The 
last term in the action suppresses all the nonzero modes. In this case, 
one can drop out the last two terms in Eq.~(\ref{SUSY action}) and put 
$Q$ to 
be a constant 
matrix \cite{Efetov83}. In this way, the universal RMT results are 
reproduced. 
For GOE,
\ifpreprintsty \else\end{multicols} 
                  	\vspace*{-3.5ex} \hspace*{0.491\hsize} 
			\begin{minipage}{0.48\hsize}$\,$\hrule
			 \end{minipage}
\fi
\begin{eqnarray}
R^{\rm o}(\omega)&=&1-\frac{\Delta^4}{\pi^4\omega^2}
\sin^2\frac{\pi\omega}{\Delta}\frac{d}{d\omega}\left[
\frac{1}{\omega}\sin \frac{\pi\omega}{\Delta}\right]
\int_1^\infty\frac{1}{t}\sin \frac{\pi\omega t}{\Delta} dt\nonumber\\
&\simeq&1-\frac{\Delta^2}{\pi^2\omega^2}-\frac{\Delta^4}{\pi^4\omega^4}
\left(\frac{3}{2}+\frac{1}{2}
\cos \frac{2\pi\omega}{\Delta}\right).
\label{zeroRo}
\end{eqnarray}
For GUE, 
\begin{equation}
R^{\rm u}(\omega)=1-\frac{\Delta^2}{\pi^2\omega^2}\sin^2\frac{\pi\omega}{\Delta}
=1-\frac{\Delta^2}{2\pi^2\omega^2}\left(1-\cos \frac{2\pi\omega}{\Delta}\right).
\label{zeroRu}
\end{equation}
In the second line of Eq.~(\ref{zeroRo}), 
we take the limit $\omega\gg\Delta$. Actually, for large $\omega$ 
$R(\omega)$ can be asymptotically expressed as
\begin{equation}
R(\omega)=R_p(\omega)+R_{np}(\omega),
\label{rfinal1}
\end{equation}
\begin{equation}
R_p(\omega)=1+\sum_n\left(\frac{\Delta}{\omega}\right)^n C_n,
\label{rfinal2}
\end{equation}
\begin{equation}
R_{np}(\omega)=
\cos \frac{2\pi\omega}{\Delta}
\sum_n\left(\frac{\Delta}{\omega}\right)^nD_n+
\sin \frac{2\pi\omega}{\Delta}
\sum_n\left(\frac{\Delta}{\omega}\right)^nE_n,
\label{rfinal3}
\end{equation}
\ifpreprintsty \else 
                  	\vspace*{-3.5ex} 
			\begin{minipage}{0.48\hsize}$\,$\hrule
			\end{minipage}\vspace*{1.5ex}
                  \begin{multicols}{2}
                   \fi\noindent
where $C_n,D_n$ and $E_n$ are universal numerical constants. 
$R(\omega)$ consists of two parts: One [denoted 
by $R_p(\omega)$] is perturbative in $\Delta/\omega$ and the other [denoted by 
$R_{np}(\omega)$] is 
nonperturbative. It is important that expanding 
$R_p(\omega)$ in 
terms of $\Delta/\omega$ is a result of making the saddle point approximation to Eqs.~(\ref{R}) and 
(\ref{SUSY action}) (see Sec.~\ref{sec3}). That is, we expand $Q$ near 
$\Lambda$ (i.e., $T$ 
near $1$) and take into account the fluctuations of $T$ perturbatively. 
In 
particular, terms like $\left(\Delta/\omega\right)^{n}, n>2$ come from perturbative 
corrections to the 
Gaussian approximation in Eq.~(\ref{R}). Actually in the zero mode 
nonlinear 
$\sigma$ model, the action depends only on the parameter 
$\omega/\Delta$. As a result, the expansion of $R(\omega)$, Eqs.~(\ref{rfinal1})-(\ref{rfinal3}) has no other parameter 
dependence. The reason is that in the disordered 
case, the Hikami box \cite{Hikami80} of zero modes has dispersion only 
at $\omega\sim 
\tau_{tr}^{-1}\gg E_{Th}$, which are unimportant in the universal 
region $\omega\ll E_{Th}$. 
For $\omega\gtrsim E_{Th}$, nonzero diffusive modes are important and $R_p(\omega)$ 
becomes nonuniversal \cite{AS86}. In the disordered case, the weak localization correction to 
nonuniversal $R_p(\omega)$ \cite{AS86} is studied in 
Ref.~\onlinecite{SLA98}.

\subsection{The Ehrenfest time $t_E$}
\label{sec2.4} 

As in the diffusive case \cite{Efetov83}, one may try to use the zero mode only and thereby drop out the free Liouville 
term as well as the regularizer in Eq.~(\ref{R}) in the region 
$\omega\ll 
t_{erg}^{-1}$ to 
recover universal RMT results. 
However, this procedure may miss important physics as we will present in 
this paper. To favor uniform $Q$, a necessary condition is that the 
regularizer becomes comparable to the Liouville term, which is of the 
order of $\sim v_F/L$. As the initial nonuniform $Q$ relaxes to the 
zero mode, 
the regularizer gets increasing $\sim \frac{1}{\tau_q}e^{\lambda t}$ 
due to the Lyapunov instability. The two terms become comparable at $t\sim t_E=\lambda^{-1}|\ln (v_F\tau_q/L)|$. Hence we 
conclude that only in the region $\omega\ll t_E^{-1}$, $Q$ can 
be put to be a constant matrix in the supersymmetric action to reproduce 
the universal RMT results. 

It is important that as usual, in the region $\omega\ll t_E^{-1}$, the 
diffusons become self-averaging in an individual quantum dot. The 
reason is that the time scale for the deterministic chaos to become 
random is $t_{erg}$. However, the time for two identical trajectories (up to the Heisenberg uncertainty) 
to decouple (i.e., become independent of each other) is much 
longer, as we will show in Sec.~\ref{sec5}, is of the order of $t_E$. 
Thus diffusons with the same coordinates are strongly coupled in the 
region 
$t_E^{-1}\sim\omega\ll t_{erg}^{-1}$. The Hikami box thereby acquires 
an additional dispersion at $\omega\sim t_E^{-1}$ and crossovers to the 
disordered limit at $\omega\ll t_E^{-1}$. As 
we mentioned earlier, for GOE, the appearance of $\left(\Delta/\omega\right)^4$ term is a 
pure quantum effect. Indeed, it comes from the coupling 
between two zero mode diffusons. More strictly speaking, it comes from the 
coupling between a diffuson and a cooperon through Hikami box.
 In this way, it is appropriate to identify such a term as the weak localization correction to 
$R_p(\omega)$. For GUE, the exact truncation at $\left(\Delta/\omega\right)^2$ term by 
no means implies the invalidity of the perturbation theory, but is due 
to the exact cancellation of the weak localization correction to $R_p(\omega)$ arising from Hikami boxes with $4$-point and 
$6$-point vertex, respectively. In Sec.~\ref{Eh_GUE_P} and \ref{Ehr_GUE_osc}, we 
will see that the $\left(\Delta/\omega\right)^4$ term shows up associating with the 
Ehrenfest oscillation in chaotic quantum dots. 

In this paper, we will study $R(\omega)$ in the region $\Delta\ll
\omega \ll t_{erg}^{-1}$. For typical quantum dots, there is only 
one macroscopic scale $L$. The Lyapunov exponent $\lambda$ is of the 
same order of $v_F/L
\sim t_{erg}^{-1}$. Thus in the crossover from semiclassical to quantum picture, the Ehrenfest time 
effectively opens a new region $\Delta\ll t_E^{-1}\sim\omega\ll 
t_{erg}^
{-1}$.

\subsection{Classical limit $\tau_q\rightarrow \infty$}
\label{sec2.3}

It is important that the regularizer introduces couplings between 
different primitive 
periodic orbits. In the absence of diffractions or quantum impurities, 
$\tau_q$ 
goes to the infinity. In such a limit, Kogan and Efetov proved that 
Eqs.~(\ref{R}) and (\ref{SUSY action}) are compatible with the 
Gutzwiller 
trace formula \cite{KE02}. Furthermore one may apply the HOA sum rule 
to 
reproduce the leading perturbative term in $R_p(\omega)$, which is of 
the order 
of $1/\omega^2$ [see Eqs.~(\ref{zeroRo}) and (\ref{zeroRu})]. 
It is important that the constant term $1$ (so-called Weyl term) in $R_p(\omega)$, is pure classical 
which characterizes the phase space volume of the energy shell 
\cite{Gut90}. So is the $\omega^{-2}$ term, which arises from the 
interference between two identical primitive periodic classical orbits 
(diagonal contribution) 
\cite{Ber85,BK96}. 
Recently, by taking into account the off-diagonal contributions, an 
oscillation 
similar to those in Eqs.~(\ref{zeroRo}) and (\ref{zeroRu}) was found 
\cite{BK96}.

Thus, we conclude that in the classical limit $\tau_q\rightarrow 
\infty$, Eqs.~(\ref{R}) and (\ref{SUSY action}) give the leading 
perturbative and nonperturbative terms. We obtain the same results by making the 
Gaussian approximation to Eqs.~(\ref{R}) 
and (\ref{SUSY action}). This is not surprising. We argue that to pass 
from the Gutzwiller formula 
to the leading $1/\omega^2$ term, the only condition is the 
ergodicity, hence it is reasonable to expect that the saddle point 
approximation 
remains good enough in the region $\Delta\ll \omega \ll t_{erg}^{-1}$. 
Actually, 
such a term results from the free motion of diffusons. It is worth 
noticing that so far the next to leading term $\sim \frac{1}{\omega^4}$ 
in 
$R_p(\omega)$, as one may expect from the universal RMT results 
[see Eq.~(\ref{zeroRo})], has not yet been identified in the framework of 
Gutzwiller formula. As we discussed in Sec.~\ref{sec2.4}, it is a pure 
quantum effect (weak localization correction).

\section{The Perturbative $R(\omega)$ In Systems With Time-Reversal Symmetry}
\label{sec3}

For systems with time-reversal symmetry, the weak localization correction to $R_p(\omega)$ in chaotic quantum dots was first 
studied by Aleiner 
and Larkin
\cite{AL97}. Furthermore, we show in this subsection that such a
correction can be written as the second derivative of the free function at the two loop levels. 
This is important
because it suggests that the result here, in principle, is 
possible to be
reproduced by using the replica technique \cite{SLA98,KM00}.

\subsection{Weak localization corrections to the perturbative $R(\omega)$}
\label{sec3.1}

From Eqs.~(\ref{R}), (\ref{pre_expan1}) and (\ref{intOrt}), we can find 
that $R_p(\omega)$ in the order of $(\Delta/\omega)^3$ is
\begin{figure}
\begin{center}
\leavevmode
\epsfxsize=7cm
\epsfbox{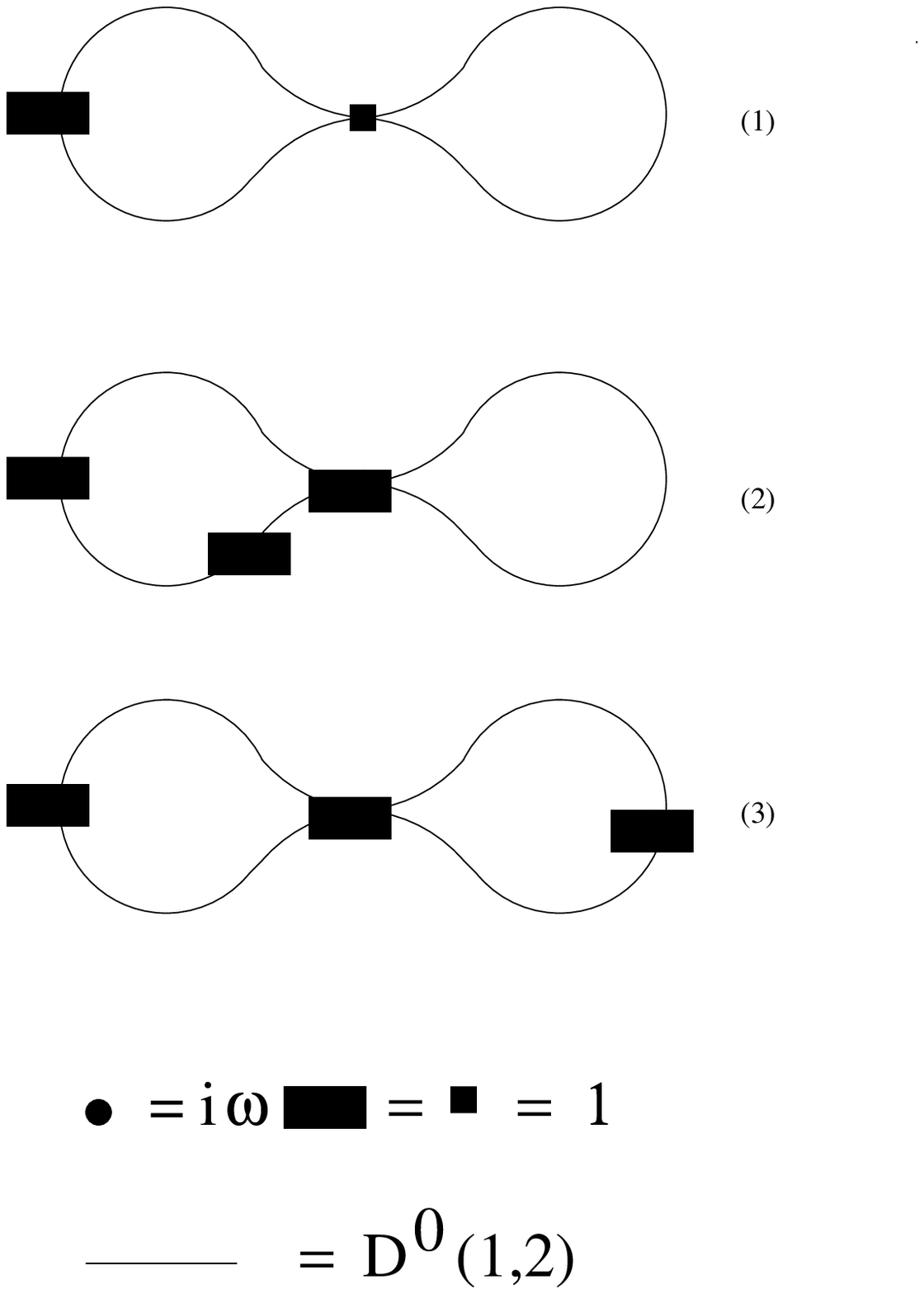}
\end{center}
\caption{ 
The diagrams contribute to $R_p(\omega)$ in the order of 
$(\Delta/\omega)^3$ in systems with time-reversal symmetry. The cross and the square, 
$\sim {\rm Str} \left[kP^2\right]$ and $\sim {\rm Str}\left[kP^4\right]$, respectively, come from the expansion in the prefactor, see Eq.~(\ref{pre_expan1}). 
The dot, $\sim i\omega{\rm Str}\left[P^4\right]$, comes from the effective interaction, see Eq.~(\ref{intOrt}).
In the figure, we put $\Delta =1$.
}
\label{fig1}
\end{figure}
\ifpreprintsty \else\end{multicols} 
                  	\vspace*{-3.5ex} \hspace*{0.491\hsize} 
			\begin{minipage}{0.48\hsize}$\,$\hrule
			 \end{minipage}
\fi
\begin{eqnarray}
\Delta R^{\rm o}_{p}(\omega) &=&
 -\frac{1}{8}{\rm Re}\int DP e^{-S^0}\int d1 {\rm 
STr}\left[kP^4(1)\right]
\int d2 {\rm STr}\left[kP^2(2)\right]-
\nonumber\\
&&\frac{1}{16}{\rm Re}\int DP e^{-S^0}
\int d1 {\rm STr}\left[kP^2(1)\right]\int d2 {\rm 
STr}\left[kP^2(2)\right]
\cdot\left(\frac{1}{2}i\pi\nu\omega\right)\int d3 {\rm 
STr}\left[P^4(3)\right],
\label{RpOrt}
\end{eqnarray}
where we keep the term with $j=1$ in the effective interaction $S^{int.}$.

To calculate Eq.~(\ref{RpOrt}), we need to use the Wick's 
theorem, which implies that any even order moments of the Gaussian integral can be factorized 
as the product of the moment $P^2$ of the Gaussian integral.  The contraction rules are \cite{AL97}:
\begin{eqnarray}
2\pi\nu P\!\!\overbrace{\ (1) M\ }\!\!P(2)&=&
 {\cal D}^0(\bar{1}, \bar{2})\Lambda_\parallel^+
 {\rm STr} \left[M\Lambda_\parallel^- \right]+
 {\cal D}^0(1, 2)\Lambda_\parallel^-
 {\rm STr}\left[ M\Lambda_\parallel^+\right]+
 {\cal D}^0(1, \bar{2})
 \Lambda_\parallel^- \bar{M}\Lambda_\parallel^+ +
 {\cal D}^0(\bar{1}, {2})
 \Lambda_\parallel^+ \bar{M}\Lambda_\parallel^-
\label{contractions_ort1}
\end{eqnarray}
and
\begin{eqnarray}
 2\pi\nu{\rm STr}[MP\!\!\overbrace{\ (1)]
{\rm STr}[N\!\!}P(2)]=
{\rm STr}\left[\left({\cal D}^0(1,2)M-{\cal 
D}^0(\bar{1},2)\bar{M}\right)
\Lambda_\parallel^-N\Lambda_\parallel^++
\left({\cal D}^0(\bar{1},\bar{2})M-{\cal D}^0(1,\bar{2})\bar{M}\right)
\Lambda_\parallel^+N\Lambda_\parallel^-
\right].
\label{contractions_ort2}
\end{eqnarray}
Here $\Lambda_\parallel^\pm =\frac{1}{2}(1\pm \Lambda)$ and $\bar M=KCM^TC^TK$.
The diffuson ${\cal D}^0(1,2)$ is the solution of the regularized Liouville (Perron-Frobenius) equation:
\begin{equation}
(-i\omega+{\hat {\cal L}}_{R,1}){\cal D}^0(1,2)=2\pi\delta(1-2).
\label{diffuson}
\end{equation}
${\cal D}^0(1,2;t)$ is the conditional probability for the particle initially at $2$ to appear at $1$. 
At the limit $\tau_q\rightarrow\infty$, ${\cal D}^0(1,2;t)=\delta({\bf R}_1-{\bf R}_2(t))\delta({\bf n}_1-{\bf n}_2(t))$ with 
$({\bf R}_2(t),{\bf n}_2(t))$ the classical trajectory starting from 
$({\bf R}_2(0),{\bf n}_2(0))=({\bf R}_2,{\bf n}_2)$.

By using Eqs.~(\ref{contractions_ort1}), (\ref{contractions_ort2}) and the following relation:
\begin{equation}
\frac{\partial}{\partial(i\omega)}{\cal D}^0(1,3)=
\int d2 {\cal D}^0(1,2){\cal D}^0(2,3),
\label{Drelation}
\end{equation} we can calculate the first term in Eq.~(\ref{RpOrt}) as  
\begin{eqnarray}
I_1&=& 2\left(\frac{\Delta}{\pi}\right)^3\int d1d2
\langle{\cal D}^0(1,2){\cal D}^0(2,{\bar 1}){\cal D}^0({\bar 
1},1)\rangle
\nonumber\\
&=&
-2\left(\frac{\Delta}{\pi}\right)^3 \int 
d1\langle\left[\frac{\partial}{\partial(-i\omega)}{\cal 
D}^0(1,\bar{1})\right]
{\cal D}^0(\bar{1},1)\rangle.
\label{dort1}
\end{eqnarray}
This gives the contraction shown in Fig.~\ref{fig1} (1). 
Then using the contraction rules, Eqs.~(\ref{contractions_ort1}) and (\ref{contractions_ort2}) we can find the second term consists of two parts: 
\begin{eqnarray}
I_2&=& 2\left(\frac{\Delta}{\pi}\right)^3i\omega\int d1d2d3
\langle{\cal D}^0(1,2){\cal D}^0(2,3)
{\cal D}^0(\bar{3},1){\cal D}^0(3,\bar{3})\rangle
\nonumber\\
&=&
-\left(\frac{\Delta}{\pi}\right)^3(-i\omega)
\int d3\langle\left[\frac{\partial^2}{\partial(-i\omega)^2}{\cal 
D}^0(\bar{3},3)\right]{\cal D}^0(3,\bar{3})\rangle,
\label{dRort2}
\end{eqnarray}
and
\begin{eqnarray}
I_3&=& -\left(\frac{\Delta}{\pi}\right)^3(-i\omega)\int d1d2d3
\langle{\cal D}^0(\bar{3},1){\cal D}^0(1,3)
{\cal D}^0(3,2){\cal D}^0(2,\bar{3})\rangle
\nonumber\\
&=&
-\left(\frac{\Delta}{\pi}\right)^3(-i\omega)\int 
d3\langle\left[\frac{\partial}{\partial(-i\omega)}{\cal 
D}^0(\bar{3},3)\right]
 \left[\frac{\partial}{\partial(-i\omega)}{\cal 
D}^0(3,\bar{3})\right]\rangle.
\label{dRort3}
\end{eqnarray}
 	\ifpreprintsty \else 
                  	\vspace*{-3.5ex} 
			\begin{minipage}{0.48\hsize}$\,$\hrule
			\end{minipage}\vspace*{1.5ex}
                  \begin{multicols}{2}
                   \fi\noindent
These correspond to the contractions, (2) and (3) in 
Fig.~\ref{fig1}, respectively. Here $\langle\cdots\rangle$ denotes the self-averaging over the 
energy spectrum. 
Adding them together, we find that Eq.~(\ref{RpOrt}) gives the weak 
localization correction to $R_p(\omega)$ as
\begin{equation}
\Delta R^{\rm o}_{p}(\omega)=
\left(\frac{\Delta}{\pi}\right)^2{\rm 
Re}\frac{\partial^2}{\partial(-i\omega)^2}F_2(\omega),
\label{DelRo}
\end{equation}
where the free energy $F_2(\omega)$ is
\begin{equation}
F_2(\omega)=i\omega
\int d1\langle{\cal D}^0(1,\bar{1}){\cal D}^0(\bar{1},1)\rangle.
\label{F2}
\end{equation}
Actually in the r.h.s. of Eq.~(\ref{F2}), one ${\cal D}^0$ comes from the diffuson and the other ${\cal D}^0$ 
comes from the cooperon \cite{DC}.

$F_2(\omega)$ is the ballistic counterpart of the free 
energy
in disordered systems, which is obtained by using the replica technique 
\cite{SLA98}. In the disordered case or in the limit $\Delta\ll\omega\ll t_E^{-1}$, 
one 
can justify that $F_2(\omega)$ is imaginary by substituting zero mode diffuson, 
i.e. ${\cal D}^0(1,2)=1/-i\omega$ into Eq.~(\ref{F2}). As a result, according to 
Eq.~(\ref{DelRo}), $\Delta R^{\rm o}_{p}(\omega)$ 
vanishes. 

\subsection{The Ehrenfest oscillations}
\label{EhrGOEper}

Now we show that
for chaotic quantum dots, in the crossover region $t_E^{-1}\sim
\omega \ll t_{erg}^{-1}$, Eqs.~(\ref{DelRo}) and (\ref{F2}) lead to the 
Ehrenfest oscillations with the period $\sim t_E^{-1}$.

We note that the diffusons ${\cal D}^0(1,{\bar 1})$ 
and ${\cal D}^0({\bar 1},1)$ are decoupled, which means
\begin{equation}
\left<{\cal D}^0(1,{\bar 1}){\cal D}^0({\bar 1},1)\right>=
\left<{\cal D}^0(1,{\bar 1})\right>\left<{\cal D}^0({\bar 1},1)\right>.
\label{2D}
\end{equation} 
Actually ${\cal D}^0(1,{\bar 1})$ stands for the probability of a trajectory initiating from $1$ and 
returning to ${\bar 1}$ in the phase space, while ${\cal 
D}^0({\bar 
1},1)$ stands for the probability of a trajectory initiating from ${\bar 1}$ and 
returning to $1$. Such two trajectories start from the same place in the real space with opposite 
directions. In this way, they are distant parts in the 
phase space. Therefore, we have Eq.~(\ref{2D}).

In Ref.~\onlinecite{AL96}, it was found that 
\begin{equation}
{\cal D}^0(1,{\bar 1})=\frac{\Gamma_2(\omega)}{-i\omega^+}.
\label{diffuson1}
\end{equation}
As a result, 
\begin{equation}
\left<{\cal D}^0(1,{\bar 1}){\cal D}^0({\bar 1},1)\right>=
\frac{\Gamma_2^2(\omega)}{(-i\omega)^2}.
\label{Hikami}
\end{equation}
From this we see that $\Gamma_2^2(\omega)$ is the effective Hikami box. In the limit 
$\omega\ll t_E^{-1}, \Gamma_2^2(\omega)\rightarrow 1$. The Hikami box becomes a constant, as in the 
diffusive disordered case \cite{Hikami80}. In the crossover region, it acquires an 
additional dispersion at $\omega\sim t_E^{-1}$.

Substituting Eqs.~(\ref{2D}) and (\ref{Hikami}) into 
Eq.~(\ref{DelRo}), 
we obtain Eq.~(\ref{Intro_1})\cite{AL97,AL97'}. Thus $R_p^{\rm o}$ acquires an Ehrenfest oscillation correction with the 
period $\sim t_E^{-1}$. Such an oscillation originates from 
the 
coupling between a diffuson and a cooperon, which is the 
reminiscence of 
Hikami box 
in the quantum disorders. In this way, essentially it is different from the 
Wigner-Dyson oscillation  In the low $\omega$ limit 
$\Delta\sim\omega\ll 
t_E^{-1}$, such 
an oscillation gradually diminishes. Actually similar corrections to 
$R_p(\omega)$ 
persist in higher order terms.

Alternatively, one may Fourier transform $R(\omega)$ as the form factor 
$K(t)=\frac{\pi}{\Delta}\int \tilde R^{\rm o}
(\omega)e^{-i\frac{\pi\omega}{\Delta} t}d\omega$ before taking the real part. 
$R(\omega)={\rm Re}\tilde R^{\rm o}(\omega)$ [cf. Eq.~(\ref{Intro_1})]. The RMT predicts 
$K(t)\simeq 2t-t^2$ at $t\ll 1$ \cite{Efetov83,Meh90}. We know that the 
linear term is pure classical (see Sec.~\ref{sec2.3}). In fact, from Eq.~(\ref{Intro_1}) we see that $K(t)=2t-t^2\theta (t-t_E)$ at $\lambda_2=0$. 
At $t<t_E$ the $t^2$ term disappears, while at $t>t_E$  it coincides with that predicted by the zero dimensional 
nonlinear $\sigma$ model \cite {Efetov83}. In Ref.~\onlinecite{SR01}, 
it is proved that in the case of $\lambda_2=0$, classical trajectories do not contribute to the 
$-t^2$ term even if the off-diagonal contributions are considered. To reproduce this 
term, the authors of Ref.~\onlinecite{SR01} introduce some correction to the 
probability of self-crossing classical trajectories.

In this section, we prove the existence of the free energy at the two loop level by using BNS$\sigma$M [see Eqs.~(\ref{DelRo}), (\ref{F2}) and (\ref{Intro_1})]. In fact, as we will see in the next section, at the three-loop level, the free energy does exist also. In this way, it is natural to expect that in general, we can write the perturbative part of $R(\omega)$ as the second order derivative of the free energy $F(\omega)$ i.e.,
\begin{equation}
R_{p}(\omega)=
\left(\frac{\Delta}{\pi}\right)^2{\rm 
Re}\frac{\partial^2}{\partial(-i\omega)^2}F(\omega). 
\label{DelRFr}
\end{equation}
Eq.~(\ref{DelRFr}) was first established for quantum disorders by using the replica technique 
\cite{SLA98}. We are not aware of any proof about it in quantum chaos.

\section{The Perturbative $R(\omega)$ In Systems With Broken Time-Reversal Symmetry}
\label{sec4}

In Sec.~\ref{sec3} we see that the weak localization correction to $R_p(\omega)$ at the two loop level originates from the coupling between a diffuson and a cooperon. In the broken time-reversal systems, the cooperon modes are suppressed by the applied magnetic field. As a result, $Q$ takes the components $k=0,3$ only in the $d$ space [see Eq.~(\ref{Pauli matrices})]. No longer exist the contraction rules Eqs.~(\ref{contractions_ort1}) and (\ref{contractions_ort2}). Instead, we can find the contraction rules below:
 \ifpreprintsty \else\end{multicols} 
                  	\vspace*{-3.5ex} \hspace*{0.491\hsize} 
			\begin{minipage}{0.48\hsize}$\,$\hrule
			 \end{minipage}
\fi
\begin{eqnarray}
4\pi\nu P\!\!\overbrace{\ (1) M\ }\!\!P(2)&=&\!
 {\cal D}^0(\bar{1}, \bar{2})\Lambda_\parallel^+
 {\rm STr} \left[M\Lambda_\parallel^- \right]+
 {\cal D}^0(1, 2)\Lambda_\parallel^-
 {\rm STr}\left[ M\Lambda_\parallel^+\right]+\nonumber\\ 
&&
 {\cal D}^0(\bar{1}, \bar{2})
 \Lambda_\parallel^+\tau_3{\rm 
STr}\left(M\Lambda_\parallel^-\tau_3\right)+
 {\cal D}^0(1,2)
 \Lambda_\parallel^-\tau_3{\rm 
STr}\left(M\Lambda_\parallel^+\tau_3\right),
\label{contractions_uni1}
\end{eqnarray}
and
\begin{equation}
2\pi\nu{\rm STr}[MP\!\!\overbrace{\ (1)]
{\rm STr}[N\!\!}P(2)]=
{\rm STr}\left[\left({\cal D}^0(1,2)M-{\cal 
D}^0(\bar{1},2)\bar{M}\right)
\Lambda_\parallel^-N\Lambda_\parallel^++
\left({\cal D}^0(\bar{1},\bar{2})M-{\cal D}^0(1,\bar{2})\bar{M}\right)
\Lambda_\parallel^+N\Lambda_\parallel^-
\right]
\label{contractions_uni2}
\end{equation}
\ifpreprintsty \else 
                  	\vspace*{-3.5ex} 
			\begin{minipage}{0.48\hsize}$\,$\hrule
			\end{minipage}\vspace*{1.5ex}
                  \begin{multicols}{2}
                   \fi\noindent
by direct calculations. Then applying the Wick's theorem and the contraction rules to Eq.~(\ref{RpOrt}), 
we find that it vanishes as expected.

In the disordered case, it was first found that the leading weak localization corrections to $R_p(\omega)$ in the broken time-reversal system are at the three loop level by using the replica technique \cite{SLA98}. To our best knowledge, the weak localization of $R_p(\omega)$ in the quantum chaotic systems with the broken time-reversal symmetry has not yet been studied. In this section, we 
prove that for the latter case, $R_p(\omega)$ can be expressed as the second derivative of the free energy at the three loop level. Moreover, we find that the $(\Delta/\omega)^4$ 
term is not exactly zero as expected by the universal Wigner-Dyson statistics. Instead, it is an Ehrenfest oscillation, which shows a crossover to its universal limit at 
$\omega\lesssim t_E^{-1}$. 

\subsection{Weak localization corrections to the perturbative $R(\omega)$}
\label{WLC_GUE}

From 
Eqs.~(\ref{R}), 
(\ref{pre_expan1}) and (\ref{intOrt}), we can find 
that $R_p(\omega)$ in the order of $(\Delta/\omega)^4$ is
\begin{equation}
\Delta R^{\rm u}_{p}(\omega)=
\Delta R^{\rm u}_{p,3b}(\omega)+
\Delta R^{\rm u}_{p,3d}(\omega).
\label{Ru3b3d}
\end{equation}
Here
\ifpreprintsty \else\end{multicols} 
                  	\vspace*{-3.5ex} \hspace*{0.491\hsize} 
			\begin{minipage}{0.48\hsize}$\,$\hrule
			 \end{minipage}
\fi
\begin{eqnarray}
\Delta R^{\rm u}_{p,3b}(\omega) &=&
\frac{1}{16}{\rm Re}\int DP e^{-S^0}\int d1 {\rm 
STr}\left[kP^2(1)\right]
\int d2{\rm STr}\left[kP^6(2)\right]+\nonumber\\
&&\frac{1}{16}{\rm Re}\int DP e^{-S^0}
\int d1 {\rm STr}\left[kP^2(1)\right]
\int d2 {\rm STr}\left[kP^2(2)\right]
\cdot\left(\frac{1}{2}i\pi\nu\omega\right)
\int d3 {\rm STr}\left[P^6(3)\right],
\label{RpUni}
\end{eqnarray}
where we keep the term with $j=2$ in the effective interaction $S^{int.}$. 
\begin{eqnarray}
\Delta R^{\rm u}_{p,3d}(\omega) &=&
\frac{1}{16}{\rm Re}\int DP e^{-S^0}\int d1 {\rm 
STr}\left[kP^4(1)\right]
\int d2{\rm STr}\left[kP^4(2)\right]+\nonumber\\
&&\frac{1}{16}{\rm Re}\int DP e^{-S^0}
\int d1 {\rm STr}\left[kP^2(1)\right]
\int d2 {\rm STr}\left[kP^4(2)\right]
\cdot\left(\frac{1}{2}i\pi\nu\omega\right)
\int d3 {\rm STr}\left[P^4(3)\right]+\nonumber\\
&&\frac{1}{32}{\rm Re}\int DP e^{-S^0}
\int d1 {\rm STr}\left[kP^2(1)\right]
\int d2 {\rm STr}\left[kP^2(2)\right]
\cdot\left(\frac{1}{2}i\pi\nu\omega\right)^2
\int d3 {\rm STr}\left[P^4(3)\right]
\int d4 {\rm STr}\left[P^4(4)\right].\nonumber\\
\label{RpUni'}
\end{eqnarray}
In Eq.~(\ref{RpUni'}), we consider up to the second order effective interaction kept 
with the term $j=1$.

First, we consider Eq.~(\ref{RpUni}). The first term in 
Eq.~(\ref{RpUni}) is found to be
\begin{eqnarray}
\Delta R^{\rm u}_{p,b1}(\omega)
&=&-\frac{3}{4}\left(\frac{\Delta}{\pi}\right)^4 \int d1d2
 \langle
{\cal D}^0(1,2){\cal D}^0(2,1){\cal D}^0(2,2){\cal D}^0(2,2)
\rangle\nonumber\\
&=&
\frac{3}{4}\left(\frac{\Delta}{\pi}\right)^4 \int d1
 \langle
 \left[\frac{\partial}{\partial(-i\omega)}{\cal D}^0(2,2)\right]
 {\cal D}^0(2,2){\cal D}^0(2,2)
 \rangle
\label{b1}
\end{eqnarray}
by using Eqs.~(\ref{contractions_uni1}) and 
(\ref{contractions_uni2}). This gives the contraction shown in (b1), 
Fig.~\ref{fig2}. 
The other term in Eq.~(\ref{RpUni}) gives 
\begin{eqnarray}
\Delta R^{\rm u}_{p,b2}(\omega)
&=& \frac{3}{4}\left(\frac{\Delta}{\pi}\right)^4(-i\omega)\int d1d2d3
\langle
{\cal D}^0(1,3){\cal D}^0(3,1){\cal D}^0(2,3){\cal D}^0(3,2){\cal 
D}^0(3,3)
\rangle
\nonumber\\
&=& \frac{3}{4}\left(\frac{\Delta}{\pi}\right)^4(-i\omega)\int d3
\langle
 \left[\frac{\partial}{\partial(-i\omega)}{\cal D}^0(3,3)\right]
 \left[\frac{\partial}{\partial(-i\omega)}{\cal D}^0(3,3)\right]
 {\cal D}^0(3,3)
 \rangle,
\label{b2}
\end{eqnarray}
and
\begin{eqnarray}
\Delta R^{\rm u}_{p,b3}(\omega)
&=&\frac{3}{4}\left(\frac{\Delta}{\pi}\right)^4(-i\omega)
\int d1d2d3
\langle
{\cal D}^0(3,1){\cal D}^0(1,2){\cal D}^0(2,3){\cal D}^0(3,3){\cal 
D}^0(3,3)
\rangle
\nonumber\\
&=&\frac{3}{8}\left(\frac{\Delta}{\pi}\right)^4(-i\omega)
\int d3\langle\left[\frac{\partial^2}{\partial(-i\omega)^2}{\cal 
D}^0(3,3)
\right]{\cal D}^0(3,3){\cal D}^0(3,3)\rangle, 
\label{b3}
\end{eqnarray}
which corresponds to two ways of contraction 
shown in (b2) and (b3) Fig.~\ref{fig2}.
Adding them together, we find the free energy $F_{3b}(\omega)$ to be [cf. Eq.~(\ref{DelRFr})]
\begin{equation}
F_{3b}(\omega)=\frac{\Delta^2}{8\pi^2}(-i\omega)
\int d1\langle{\cal D}^0(1,1){\cal D}^0(1,1){\cal D}^0(1,1)\rangle.
\label{F3b}
\end{equation}

Now we turn to consider Eq.~(\ref{RpUni'}). The first term in 
Eq.~(\ref{RpUni'}) is reduced into 
\begin{equation}
\Delta R^{\rm u}_{p,d1}(\omega)
=-\frac{1}{4}\left(\frac{\Delta}{\pi}\right)^4\int d1d2\langle
{\cal D}^0(1,2){\cal D}^0(1,2)
{\cal D}^0(2,1){\cal D}^0(2,1)\rangle
\label{d1}
\end{equation}
by using Eqs.~(\ref{contractions_uni1}) and 
(\ref{contractions_uni2}), which gives the contraction shown in (d1), 
Fig.~\ref{fig22}.
The second term in Eq.~(\ref{RpUni'}) is
\begin{eqnarray}
\Delta R^{\rm u}_{p,d2}(\omega)
&=&2\left(\frac{\Delta}{\pi}\right)^4(-i\omega)\int d1d2d3
\langle
{\cal D}^0(2,3){\cal D}^0(2,3){\cal D}^0(3,2){\cal D}^0(3,1){\cal 
D}^0(1,2)
\rangle
\nonumber\\
&=&-2\left(\frac{\Delta}{\pi}\right)^4(-i\omega)\int d2d3
\langle
{\cal D}^0(2,3){\cal D}^0(2,3)
{\cal D}^0(3,2)
\left[
\frac{\partial}{\partial(-i\omega)}{\cal D}^0(3,2)
\right]
\rangle, 
\label{d2}
\end{eqnarray}
which gives the contraction shown in 
(d2), Fig.~\ref{fig22}.
The diagram (d3) and (d4) in Fig.~\ref{fig22} come from the third term 
in 
Eq.~(\ref{RpUni'}):
\begin{eqnarray}
\Delta R^{\rm u}_{p,d3}(\omega)
&=&
-\left(\frac{\Delta}{\pi}\right)^4(-i\omega)^2\int d1d2d3d4
\langle
{\cal D}^0(3,4){\cal D}^0(3,4){\cal D}^0(4,3)
{\cal D}^0(4,2){\cal D}^0(2,1){\cal D}^0(1,3)
\rangle
\nonumber\\
&=&
-\frac{1}{2}\left(\frac{\Delta}{\pi}\right)^4(-i\omega)^2\int d3d4
\langle
{\cal D}^0(3,4){\cal D}^0(3,4){\cal D}^0(4,3)
\left[\frac{\partial ^2}{\partial(-i\omega)^2}{\cal D}^0(4,3)\right]
\rangle,
\label{d3}
\end{eqnarray}
\begin{eqnarray}
\Delta R^{\rm u}_{p,d4}(\omega)
&=&-\left(\frac{\Delta}{\pi}\right)^4(-i\omega)^2\int 
d1d2d3d4
\langle
{\cal D}^0(3,2){\cal D}^0(2,4){\cal D}^0(3,4){\cal D}^0(4,3){\cal 
D}^0(4,1){\cal D}^0(1,3)
\rangle-
\nonumber\\
&&\frac{1}{2}\left(\frac{\Delta}{\pi}\right)^4(-i\omega)^2\int 
d1d2d3d4
\langle
{\cal D}^0(3,1){\cal D}^0(1,4){\cal D}^0(4,3){\cal D}^0(3,2){\cal 
D}^0(2,4){\cal D}^0(4,3)
\rangle
\nonumber\\
&=&-\left(\frac{\Delta}{\pi}\right)^4(-i\omega)^2\int d3d4
\langle
\left[\frac{\partial}{\partial(-i\omega)}{\cal D}^0(3,4)\right]
{\cal D}^0(3,4){\cal D}^0(4,3)
\left[\frac{\partial}{\partial(-i\omega)}{\cal D}^0(4,3)\right]
\rangle-
\nonumber\\
&&\frac{1}{2}
\left(\frac{\Delta}{\pi}\right)^4(-i\omega)^2\int d3d4
\langle\left[\frac{\partial}{\partial(-i\omega)}{\cal D}^0(3,4)\right]
\left[\frac{\partial}{\partial(-i\omega)}{\cal D}^0(3,4)\right]
{\cal D}^0(4,3){\cal D}^0(4,3)\rangle.
\label{d4}
\end{eqnarray}
Collecting Eqs.~(\ref{d1})-(\ref{d4}) together, we find the free 
energy $F_{3d}(\omega)$ to be [cf. Eq.~(\ref{DelRFr})]
\begin{equation}
F_{3d}(\omega)=-\frac{\Delta^2}{8\pi^2}(-i\omega)^2\int d1d2\langle
{\cal D}^0(1,2){\cal D}^0(1,2){\cal D}^0(2,1){\cal D}^0(2,1)\rangle.
\label{F3d}
\end{equation}
\ifpreprintsty \else 
                  	\vspace*{-3.5ex} 
			\begin{minipage}{0.48\hsize}$\,$\hrule
			\end{minipage}\vspace*{1.5ex}
                  \begin{multicols}{2}
                   \fi\noindent

Hence we prove Eq.~(\ref{DelRFr}) at the three loop level. We see that Eqs.~(\ref{RpUni}) and (\ref{RpUni'}) 
give the weak localization correction to $R_p(\omega)$, up to the order 
of $(\Delta/\omega)^4$ as
\begin{equation}
\Delta R^{\rm u}_p(\omega)=
\left(\frac{\Delta}{\pi}\right)^2 {\rm Re}
\frac{\partial^2}{\partial(-i\omega)^2}
\left(
F_{3b}(\omega)+F_{3d}(\omega)
\right).
\label{DelRpu}
\end{equation}

\subsection{The Couplings of Diffusons}
\label{multi_diffusons}

In this subsection we will calculate the product of diffusons appearing in Eqs.~(\ref{F3b}) and (\ref{F3d}).

We emphasize that it is difficult to use the standard 
diagram technique \cite{AGD} 
to calculate these quantities. Instead, it is more convenient to use the semiclassical 
method. In 
this section, we will follow the techniques of Ref.~\onlinecite{AL96}. This immediately leads to the Ehrenfest 
oscillations 
in the perturbative $R(\omega)$. It is important that the average $\langle
\cdots\rangle$ is performed over the energy for fixed potentials, not 
impurity configurations.

\subsubsection{ The couplings of two diffusons}
\label{two_diffus}

To calculate Eq.~(\ref{F3d}), we notice that there is the following relation:
\begin{eqnarray}
&&\left<{\cal D}^0(1,2){\cal D}^0(1,2)
{\cal D}^0(2,1){\cal D}^0(2,1)\right>\nonumber\\
&=&
\left<{\cal D}^0(1,2){\cal D}^0(1,2)\right>
\left<{\cal D}^0(2,1){\cal D}^0(2,1)\right>
\label{4D}
\end{eqnarray} 
for similar reasons as discussions for Eq.~(\ref{2D}). Actually ${\cal 
D}^0(1,2)$ 
stands for the probability of a trajectory initiating from $1$ and 
ending at $2$ in the phase space, while ${\cal D}^0(2,1)$ stands for the 
probability of a trajectory initiating from $2$ and 
ending at $1$. Thus such two trajectories are distant 
parts and decoupled. But $\langle {\cal D}^0(1,2){\cal D}^0(1,2)\rangle$ 
\begin{figure}
\begin{center}
\leavevmode
\epsfxsize=7cm
\epsfbox{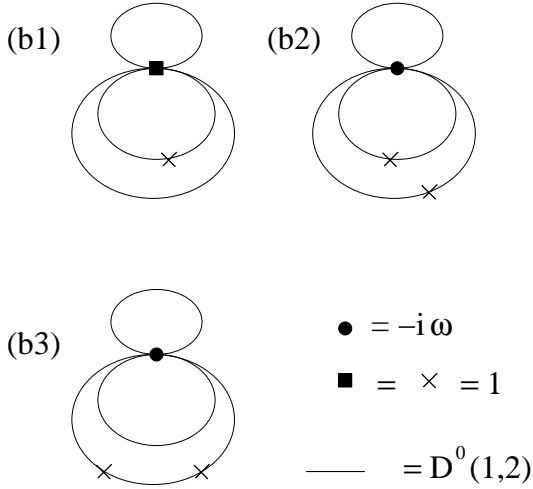}
\end{center}
\caption{ The diagrams with $6$-point vertex contribute to $R_p(\omega)$ in the order of 
$(\Delta/\omega)^4$ in systems with broken time-reversal symmetry.
The cross and the square, 
$\sim {\rm Str} \left[kP^2\right]$ and $\sim {\rm Str}\left[kP^6\right]$, respectively, come from the expansion in the prefactor, see Eq.~(\ref{pre_expan1}). 
The dot, $\sim i\omega{\rm Str}\left[P^6\right]$, comes from the effective interaction, see Eq.~(\ref{intOrt}).
In the figure, we put $\Delta =1$. 
}
\label{fig2}
\end{figure}
\noindent
can not be 
factorized at the region near $1$ or $2$, where two trajectories become 
very close to each other.

We will study the quantity $\langle {\cal 
D}^0(1_1,2) {\cal 
D}^0(1_2,2)\rangle$ instead of $\langle {\cal D}^0(1,2) 
{\cal D}^0(1,2)\rangle $ and put $1_1=1_2$ in the final answer. Here we use 
the subscripts $1$ and $2$ to denote the small deviations in the momentum direction 
and position. From Eq.~(\ref{diffuson}), we find the motion 
equation to be 

\begin{figure}
\begin{center}
\leavevmode
\epsfxsize=7cm
\epsfbox{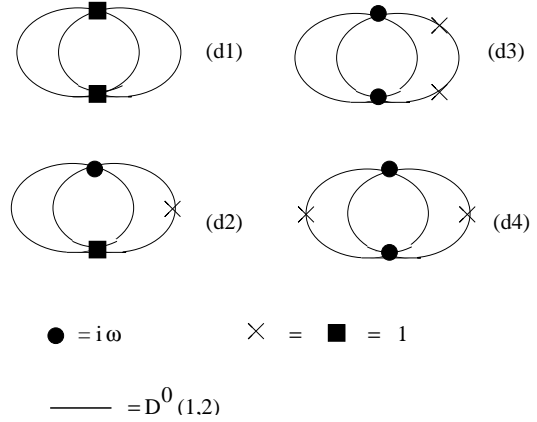}
\end{center}
\caption{ The diagrams with $4$-point vertex contribute to $R_p(\omega)$ in the order of 
$(\Delta/\omega)^4$ in systems with broken time-reversal symmetry.
The cross and the square, 
$\sim {\rm Str} \left[kP^2\right]$ and $\sim {\rm Str}\left[kP^4\right]$, respectively, come from the expansion in the prefactor, see Eq.~(\ref{pre_expan1}). 
The dot, $\sim i\omega{\rm Str}\left[P^4\right]$, comes from the effective interaction, see Eq.~(\ref{intOrt}).
In the figure, we put $\Delta =1$.
}
\label{fig22}
\end{figure}
\ifpreprintsty \else\end{multicols} 
                  	\vspace*{-3.5ex} \hspace*{0.491\hsize} 
			\begin{minipage}{0.48\hsize}$\,$\hrule
			 \end{minipage}
\fi
\begin{equation}
\left(-2i\omega+{\hat {\cal L}}_{1_1}+{\hat {\cal L}}_{1_2}-
\frac{1}{\tau_q}\frac{\partial^2}{\partial\phi_{1_1}^2}-
\frac{1}{\tau_q}\frac{\partial^2}{\partial\phi_{1_2}^2}\right)
{\cal D}^0(1_1,2) {\cal D}^0(1_2,2)
=2\pi\left[
\delta_{1_1,2}{\cal D}^0(1_2,2)+ \delta_{1_2,2}{\cal 
D}^0(1_1,2)\right].
\label{2Dmotion}
\end{equation} 
 	\ifpreprintsty \else 
                  	\vspace*{-3.5ex} 
			\begin{minipage}{0.48\hsize}$\,$\hrule
			\end{minipage}\vspace*{1.5ex}
                  \begin{multicols}{2}
                   \fi\noindent
We will use $({\bf n}_1, {\bf R}_1)$ to denote the motion of the center of 
mass while 
$(\phi_1, \rho_1)$ to denote the relative motion \cite{AL96}. The operator $\left({\hat {\cal L}}_{1_1}+{\hat {\cal L}}_{1_2}\right)$ 
is written as 
\begin{eqnarray}
{\hat {\cal L}}_1+{\hat {\cal L}}_2=
{\hat {\cal L}}_{c,1}+{\hat {\cal L}}_{r,1}
\nonumber\\
{\hat {\cal L}}_{c,1}=v_F{\bf n}_1\cdot\frac{\partial}{\partial {\bf 
R}_1}-
\nabla U({\bf R}_1)\cdot\frac{\partial}{\partial {\bf p}_1},
\nonumber\\
{\hat {\cal L}}_{r,1}=-v_F\phi_1\frac{\partial}{\partial\rho_1}+
\frac{\partial^2 U}{p_F\partial 
R_{\bot,1}^2}\rho_1\frac{\partial}{\partial\phi_1},\nonumber\\
\label{relmotion'}
\end{eqnarray} 
where $R_{\bot}$ denote the coordinate perpendicular to the momentum 
direction. For the convenience below, we further introduce the change of variables:
\begin{equation}
z_1=\ln\left[\phi_1^2+\left(\frac{\rho_1}{L}\right)^2\right]^{1/2},\qquad
\alpha_1=\arctan \frac{\phi_1 L}{\rho_1}.
\label{zalpha}
\end{equation}
At $\omega\ll\lambda$, the deterministic motion of the center of mass becomes random. 
Thus, the quantity ${\cal D}^0(1_1,2) {\cal D}^0(1_2,2)$ is self-averaging over ${\bf R}_1, {\bf n}_1$. 
In this way, we can introduce the following quantity:
\begin{eqnarray}
M(\phi_1,\rho_1;2,2)&\equiv&\langle{\cal D}^0(1_1,2) {\cal 
D}^0(1_2,2)\rangle
\nonumber\\
&=&\int \frac{d{\bf R}_1d{\bf n}_1}{2\pi A}{\cal D}^0(1_1,2) {\cal 
D}^0(1_2,2).
\label{selave}
\end{eqnarray}

By definition we imply that $|z_1|\simeq |z_2|\gg 1$. In this way, we can see that 
the particular solution due to the r.h.s. of Eq.~(\ref{2Dmotion}) is of the order of $1/\lambda\omega$ 
and depends on $z_1-z_2\simeq 0$. Therefore, we can ignore the r.h.s. of Eq.~(\ref{2Dmotion}) 
since the particular solution is much smaller than $1/\omega^2$ in the universal region: 
$\omega\ll\lambda$. Then averaging Eq.~(\ref{relmotion'}) over ${\bf n}_1, {\bf R}_1$ and 
$\alpha_1$, we obtain \cite{AL96}:
\ifpreprintsty \else\end{multicols} 
                  	\vspace*{-3.5ex} \hspace*{0.491\hsize} 
			\begin{minipage}{0.48\hsize}$\,$\hrule
			 \end{minipage}
\fi
\begin{equation}
\left[2i\omega+\lambda\frac{\partial}{\partial 
z_1}+\frac{\lambda_2}{2}
\frac{\partial^2}{\partial z_1^2}+\frac{e^{-2z_1}}{2\tau_q}
\frac{\partial}{\partial z_1}
\left(\frac{1-\gamma}{2}\frac{\partial}{\partial z_1}+\gamma\right)
\right]M(z_1;2,2)=0.
\label{C'}
\end{equation}
\ifpreprintsty \else 
                  	\vspace*{-3.5ex} 
			\begin{minipage}{0.48\hsize}$\,$\hrule
			\end{minipage}\vspace*{1.5ex}
                  \begin{multicols}{2}
                   \fi\noindent
Eq.~(\ref{C'}) exists only at the Lyapunov region, i.e., $\rho\ll L, \phi\ll 1 (z_1\ll -1)$. 
At $z_1\sim 0$, Eq.~(\ref{C'}) is not applicable and the solution of $M$ does not depend on $z_1$ any more. Since we are interested in the 
region $\omega\sim t_E^{-1}$, the coordinates $1$ and $2$ are independent from each other. 
Hence $z_1\simeq z_2$, by definition, implies that the deterministic relative motion (Lyapunov asymptotic instability) of $z_1$ 
must reach $z_1=0$ in the course of time. In this way, Eq.~(\ref{C'}) must be supplemented by 
the boundary condition $M(\omega,z_1=0)=M'$. $M'$ will be 
calculated below.

The solution of Eq.~(\ref{C'}) is
\begin{equation}
M=w_2(\omega,z_1)M',
\label{M}
\end{equation} 
where  $w_2(\omega, z_1)$ is found to be \cite{AL96}
\begin{equation}
w_2(\omega,z_1)=\exp\left[
\left(\frac{i\omega}{\lambda}-\frac{\omega^2\lambda_2}{\lambda^3}
\right)
\ln \frac{\lambda\tau_q}{\lambda\tau_q e^{2z_1}+\gamma /2}\right].
\label{W2}
\end{equation}
up to the logarithmic accuracy. $\gamma\lesssim 1$ is a numerical factor. Any other form of 
the 
regularizer with the form of the second order elliptic operator changes 
the value of $\gamma$ only. 
$t_E$ is the Ehrenfest time: 
\begin{equation}
t_E=\frac{1}{\lambda}|\ln \lambda\tau_q|.
\label{tE'}
\end{equation}
In the limit $z_1\rightarrow 
-\infty$, $M$ 
becomes 
\begin{equation}
M=\Gamma_2(\omega)M'.
\label{MM}
\end{equation}
At the classical limit $\tau_q\rightarrow\infty$, $w_2\rightarrow 0$ as $z_1\rightarrow -\infty$. 
But this is not the case for finite $\tau_q$. As we see from Eq.~(\ref{W2}), the classical 
solution $w_2(\omega, z_1)$ ($\tau_q=\infty$) thereby acquires a lower cutoff at $z_1=-\frac{1}{2}\ln \lambda \tau_q$. 
That is, we may regard $w_2(\omega,-\infty)$ as $w_2(\omega,-\frac{1}{2}\ln \lambda \tau_q)$.

To find $M'$, we repeat the procedure of deriving Eq.~(\ref{2Dmotion}) 
and 
study a more general quantity: $U(\omega;{\bar 2}_1,{\bar 2}_2), M'=U(\omega;\bar 2,\bar 2)$. 
Similarly to Eq.~(\ref{C'}), $U(z_{\bar 2};\omega)$ is the solution of the equation:
\ifpreprintsty \else\end{multicols} 
                  	\vspace*{-3.5ex} \hspace*{0.491\hsize} 
			\begin{minipage}{0.48\hsize}$\,$\hrule
			 \end{minipage}
\fi
\begin{equation}
\left[2i\omega+\lambda\frac{\partial}{\partial 
z_{\bar 2}}+\frac{\lambda_2}{2}
\frac{\partial^2}{\partial z_{\bar 2}^2}+\frac{e^{-2z_{\bar 2}}}{2\tau_q}
\frac{\partial}{\partial z_{\bar 2}}
\left(\frac{1-\gamma}{2}\frac{\partial}{\partial z_{\bar 2}}+\gamma\right)
\right]U(z_{\bar 2};\omega)=0,
\label{M2'eq}
\end{equation}
\ifpreprintsty \else 
                  	\vspace*{-3.5ex} 
			\begin{minipage}{0.48\hsize}$\,$\hrule
			\end{minipage}\vspace*{1.5ex}
                  \begin{multicols}{2}
                   \fi\noindent
where $z_{\bar 2}=z_2$ characterizes the separation of two nearby trajectories centered at $\bar 2$. 
We drop out the r.h.s. for the same reasons as deriving 
Eq.~(\ref{C'}). 
To find the solution of Eq.~(\ref{M2'eq}). We repeat the 
procedure of Eqs.~(\ref{M}) to (\ref{MM}), except that the boundary 
condition is 
replaced by
\begin{equation}
U(z_{\bar 2}=0,\omega)=\frac{1}{(-i\omega)^2}.
\label{MBC}
\end{equation} 
Eq.~(\ref{MBC}) means that if the deviations $\rho_1\sim\rho_2\sim L (z_1=z_2=0)$, then the two trajectories, connecting $1_1$ and $2_1$, $1_2$ and $2_2$, respectively, are independent from each other. Thus we have $\langle {\cal D}^0{\cal D}^0\rangle=\langle {\cal D}^0\rangle ^2$. 
Similarly, we have 
\begin{equation}
M'=U(z_2=-\infty)=\Gamma_2(\omega) \frac{1}{(-i\omega)^2}.
\label{M'}
\end{equation} 
Substituting Eq.~(\ref{M'}) into 
Eq.~(\ref{MM}), we find that
\begin{equation}
\langle {\cal D}^0(1,2) {\cal D}^0(1,2)\rangle\equiv 
M=\frac{\Gamma_2^2(\omega)}{(-i\omega)^2}.
\label{Msolution}
\end{equation} 
At $\omega\ll \lambda$, 
\begin{equation}
\langle {\cal D}^0(1,1) {\cal D}^0(1,1)\rangle=
\langle {\cal D}^0(1,2) {\cal D}^0(1,2)\rangle=
\frac{\Gamma_2^2(\omega)}{(-i\omega)^2}.
\label{MMsolution}
\end{equation}

We would like to mention that in the case of long-ranged disorders with $\tau_q=\infty$), 
an equation similar to Eq.~(\ref{C'}) was found by using the path integral technique \cite{GM02}. 
In fact, in this case, $\lambda$ and $\lambda_2$ were calculated [see Appendix of Ref.~\onlinecite{AL96}]. 
It is found that $\lambda_2\simeq 4\lambda$.

\subsubsection{The couplings of three diffusons}
\label{three_diffus}

Now we turn to calculate Eq.~(\ref{F3b}). First, we note that although the coordinates in 
${\cal D}^0(1,1)$ coincide with each other, 
they are not related to each other at the region $\omega\ll\lambda$, 
because the trajectory travels for a long time before returning to the 
same position in the phase space. Thus, to calculate $\langle{\cal D}^0(1,1){\cal D}^0(1,1)
{\cal D}^0(1,1)\rangle$ in
Eq.~(\ref{F3b}), we will calculate a more general quantity, say 
$\langle{\cal D}^0(1,2){\cal D}^0(1,2){\cal D}^0(1,2)\rangle$.

We 
proceed along the 
line of calculating two coupled diffusons. Similarly to Eq.~(\ref{2Dmotion}), we have 
the following equation:
 \ifpreprintsty \else\end{multicols} 
                  	\vspace*{-3.5ex} \hspace*{0.491\hsize} 
			\begin{minipage}{0.48\hsize}$\,$\hrule
			 \end{minipage}
\fi
\begin{eqnarray}
\left(-3i\omega+{\hat {\cal L}}_{1_1}+{\hat {\cal L}}_{1_2}+
{\hat {\cal L}}_{1_3}-
\frac{1}{\tau_q}\frac{\partial^2}{\partial\phi_{1_1}^2}-
\frac{1}{\tau_q}\frac{\partial^2}{\partial\phi_{1_2}^2}-
\frac{1}{\tau_q}\frac{\partial^2}{\partial\phi_{1_3}^2}\right)
{\cal D}^0(1_1,2) {\cal D}^0(1_2,2){\cal D}^0(1_3,2)\nonumber\\
=2\pi\left[
\delta_{1_1,2}{\cal D}^0(1_2,2){\cal D}^0(1_3,2)+
\delta_{1_2,2}{\cal D}^0(1_1,2){\cal D}^0(1_3,2)+
\delta_{1_3,2}{\cal D}^0(1_1,2){\cal D}^0(1_2,2)\right].
\label{3Dmotion}
\end{eqnarray} 
Then we can separate the motion into the part of the center of 
mass and the relative part by introducing the change of variables:
\begin{eqnarray}
&&{\bf R}_1=\frac{{\bf R}_{1_1}+{\bf R}_{1_2}+{\bf R}_{1_3}}{3},\qquad
\phi_1=\frac{\phi_{1_1}+\phi_{1_2}+\phi_{1_3}}{3},\nonumber\\
&&{\bf r}_{1_1}={\bf R}_{1_1}-{\bf R}_1,\qquad
\theta_1=\phi_{1_1}-\phi_1,\nonumber\\
&&{\bf r}_{1_2}={\bf R}_{1_2}-{\bf R}_1,\qquad
\theta_2=\phi_{1_2}-\phi_1.
\label{R1R2R3}
\end{eqnarray}
Here $(\phi_1,{\bf R}_1)$ denotes the coordinates of the center of mass. 
The relative motion is described by the set of variables: 
$({\bf r}_{1_1},{\bf r}_{1_2};\theta_1,\theta_2)$. The transverse components of 
${\bf r}_{1_1}$ and ${\bf r}_{1_2}$ are denoted as $\rho_1$ and  $\rho_2$. Under 
such a change of variables, we rewrite Eq.~(\ref{3Dmotion}) as 
\begin{eqnarray}
&&\left[-3i\omega+{\hat {\cal L}}_{{\bf R}_1}+
{\hat {\cal L}}_{{\bf r}_{1_1}}+
{\hat {\cal L}}_{{\bf r}_{1_2}}-
\frac{1}{3\tau_q}\frac{\partial^2}{\partial\phi_1^2}-\frac{2}{3}\frac{1}{\tau_q}
\left(\frac{\partial^2}{\partial\theta_1^2}+
\frac{\partial^2}{\partial\theta_2^2}-
\frac{\partial^2}{\partial\theta_1\partial\theta_2}
\right)
\right]{\cal D}^0(1_1,2) {\cal D}^0(1_2,2){\cal D}^0(1_3,2)=0,\nonumber\\
&&{\hat {\cal L}}_{{\bf R}_1}=v_F{\bf n}_1\cdot\frac{\partial}{\partial {\bf R}_1}-
\nabla U({\bf R}_1)\cdot\frac{\partial}{\partial {\bf p}_1},
\nonumber\\
&&{\hat {\cal L}}_{{\bf r}_{1_1}}=
-v_F\theta_1\frac{\partial}{\partial\rho_1}+
\frac{\partial^2 U}{p_F\partial 
R_{\bot,1}^2}\rho_1\frac{\partial}{\partial\theta_1},\nonumber\\
&&{\hat {\cal L}}_{{\bf r}_{1_2}}=
-v_F\theta_2\frac{\partial}{\partial\rho_2}+
\frac{\partial^2 U}{p_F\partial 
R_{\bot,1}^2}\rho_2\frac{\partial}{\partial\theta_2}.
\label{3Dmotion'}
\end{eqnarray}
\ifpreprintsty \else 
                  	\vspace*{-3.5ex} 
			\begin{minipage}{0.48\hsize}$\,$\hrule
			\end{minipage}\vspace*{1.5ex}
                  \begin{multicols}{2}
                   \fi
\noindent
Here we expand $\theta, \rho$ up to the first order.
We again drop out the r.h.s. of Eq.~(\ref{3Dmotion}).
In Eq.~(\ref{3Dmotion'}), the form of the regularizer has changed compared to Eq.~(\ref{2Dmotion}). 
However, as we discussed in Sec.~\ref{sec2.1}, such a change accounts 
only for the value of 
$\tau_q$ under logarithm. Consequently, up to 
the logarithmic accuracy, $t_E$ remains unchanged. In view of this, we ignore the regularizer in the 
discussions below. To solve Eq.~(\ref{3Dmotion'}), we will proceed along the line of calculating two coupled diffusons. Here we will show the main result and give the details in Appendix~\ref{Deriva_DDD}. Eq.~(\ref{3Dmotion'}) can be 
reduced as 
\begin{eqnarray}
&&\left(3i\omega+\lambda\frac{\partial}{\partial x}+\frac{\lambda_2}{2}
\frac{\partial^2}{\partial x^2}
\right)N(x;2,2)=0,\nonumber\\
&&x=\frac{1}{2}\left(
\ln \sqrt{\theta_1^2+\left(\frac{\rho_1}{L}\right)^2}+
\ln \sqrt{\theta_2^2+\left(\frac{\rho_2}{L}\right)^2}
\right),
\label{DDD}
\end{eqnarray}
where $N(x;2,2)$ is the self-averaging of \\
\noindent 
${\cal D}^0(1_1,2) {\cal D}^0(1_2,2){\cal D}^0(1_3,2)$ over the center of mass ${\bf R}_1,{\bf n}_1$:
\begin{eqnarray}
&&N(x;2,2)=
N(\theta,\rho;2,2)
\nonumber\\
&\equiv&\langle{\cal D}^0(1_1,2) {\cal 
D}^0(1_2,2){\cal D}^0(1_3,2)\rangle\nonumber\\
&=&\int \frac{d{\bf R}_1d{\bf n}_1}{2\pi A}{\cal D}^0(1_1,2) {\cal 
D}^0(1_2,2){\cal D}^0(1_3,2). 
\label{Ndef}
\end{eqnarray}

The similar equation was proposed in Ref.~\onlinecite{VL02}. It is 
important that here $\lambda$ and $\lambda_2$ stay exactly the same as 
those appearing 
in $\Gamma_2(\omega)$ (see Appendix~\ref{Deriva_DDD}). Compared to 
Eq.~(\ref{C'}),  Eq.~(\ref{DDD}) differs from it 
in the frequency ($\omega\rightarrow \frac{3}{2}\omega$) and the 
variable ($z\rightarrow x$). 
Therefore, in the limit $\tau_q\rightarrow\infty$, the homogeneous solution of Eq.~(\ref{DDD}), say 
$w_3(x;\omega)$ is 
\begin{equation}
w_3(x;\omega)=w_2(x;\frac{3}{2}\omega)
\label{w3}
\end{equation} 
supplemented by the boundary condition, which is $w_3(x=0;\omega)=1$. 
Up to the logarithmic accuracy, we obtain: 
\begin{eqnarray}
\Gamma_3(\omega)&\equiv&\lim_{x\rightarrow -\infty}
w_3(x;\omega)=w_3(-\lambda^{-1}\ln \lambda \tau_q;\omega)
\nonumber\\
&=&\exp \left(
\frac{3}{2}i\omega t_E-\frac{9}{4}\frac{\omega^2\lambda_2 
t_E}{\lambda^2}
\right).
\label{Gamma3def}
\end{eqnarray} 
Then we follow the procedure of Eqs.~(\ref{M})-(\ref{Msolution}) 
to get:  
\begin{equation}
\langle{\cal D}^0(1,2) {\cal D}^0(1,2){\cal 
D}^0(1,2)\rangle= 
N(x\rightarrow -\infty,\omega)=\frac{\Gamma_3^2(\omega)}{(-i\omega)^3}.
\label{Nsolution}
\end{equation}

In fact, we have a general expression for $n-$coupled diffusons, 
say $\Gamma_n(\omega)$. One is able to prove that $\Gamma_n(\omega)$ is 
also 
oscillating: $\sim \exp (in\omega t_E/2)$ in the same spirit, 
which can be obtained in an alternative method, i.e., averaging the 
classical solution 
($\tau_q\rightarrow \infty$) over a minimal Gaussian wave packet 
\cite{VL02}. 
It is in order to make the following observations: For 
$\Gamma_n(\omega)$, 
the oscillating part does not depend on the particular choice of the 
regularizer 
except that $t_E$ depends on $\tau_q$ parametrically.

\subsection{The Ehrenfest oscillations in $R_p^{\rm u}(\omega)$}
\label{Eh_GUE_P}

$F_{3b}(\omega)$ and $F_{3d}(\omega)$ are the ballistic counterparts of 
the zero mode approximation of the 
free energy in disordered systems. In the disordered limit or $\Delta\ll\omega\ll t_E^{-1}$, one 
can verify that $F_{3b}(\omega)$ and $F_{3d}(\omega)$ cancel each 
other and Eq.~(\ref{DelRpu}) gives vanishing result for $\Delta R^{\rm 
u}_p(\omega)$ by 
putting ${\cal D}^0(1,2)$ to be zero mode diffuson, i.e. ${\cal 
D}^0(1,2)=
1/-i\omega$. However, as we will see below, it 
is not so for chaotic quantum dots in the crossover region $t_E^{-1}\sim
\omega \ll t_{erg}^{-1}$.
Instead, Eqs.~(\ref{F3b}), (\ref{F3d}) and (\ref{DelRpu})
lead to the Ehrenfest oscillation with the period $\sim t_E^{-1}$.

Substituting 
Eq.~(\ref{Nsolution}) into Eq.~(\ref{F3b}), we find that
\begin{equation}
F_{3b}(\omega)=\frac{\Delta^2}{8\pi^2}\frac{\Gamma_3^2(\omega)}{(-i\omega)^2}.
\label{f3b}
\end{equation}
Taking 
Eq.~(\ref{4D}) 
into account, we find
\begin{equation}
F_{3d}(\omega)=-\frac{\Delta^2}{8\pi^2}\frac{\Gamma_2^4(\omega)}{(-i\omega)^2}.
\label{f3d}
\end{equation}
Therefore, we obtain Eq.~(\ref{RuniEO}). We see that $R_p^{\rm u}(\omega)$ acquires an oscillation correction at $\omega t_E\sim 1$ with the period $\sim t_E^{-1}$.

We should emphasize that the appearance of the Ehrenfest 
oscillations, 
$\sim e^{in\omega t_E}, n=1,2,3,\cdots$ do not depend on a particular 
form of 
the regularizer, although $t_E$ does. This can be 
understood in the following way. In the Lyapunov region, 
$n$ coupled diffusons (Hikami box with $2n$-point vertex) contribute 
to $R(\omega)$ (in the disordered limit) an additional factor $\sim 
\left(G^R_{\epsilon+\frac{\omega}{2}}G^A_{\epsilon-\frac{\omega}{2}}\right)^n$. 
In the semiclassics, $G^R_\epsilon\sim e^{i\epsilon l/v_F}$ and 
$G^A_\epsilon\sim e^{-i\epsilon l/v_F}$with $l$ being the length of the 
trajectory piece. Up to the logarithmic accuracy, the Lyapunov regions 
have 
the same length $v_Ft_E$. In this way, 
$\left(G^R_{\epsilon+\frac{\omega}{2}}G^A_{\epsilon-\frac{\omega}{2}}
\right)^n\sim e^{in\omega t_E}$.
In general, $R_p(\omega)$ can be expressed as
\begin{eqnarray}
R_p(\omega)&=&1+\sum_n\left(\frac{\Delta}{\omega}\right)^nC_n(\omega 
t_E)
\nonumber\\
&=&1+\frac{\Delta^2}{\pi^2} \frac{\partial^2}{\partial\omega^2}F(\omega),
\nonumber\\
F(\omega)&=&\sum_L\left(
\frac{\Delta}{-i\omega}
\right)^{L-1}
\sum_{V,S}\delta_{L+V-S,1}
\tilde C_S(\omega t_E),
\label{RpF}
\end{eqnarray}
where $L,V$ and $S$ characterize the topology of the diagram of 
the free energy, from which $R_p(\omega)$ is obtained. $L,S$ and $V$ 
are 
the number of loops, sides (diffusons) and vertices (Hikami boxes) 
respectively. The constraint comes 
from the well known Euler theorem in the topology \cite{topo}. For a 
topological diagram of the free 
energy with $S$ sides (diffusons), we have 
$\left(G^R_{\epsilon+\frac{\omega}{2}}G^A_{\epsilon-\frac{\omega}{2}}\right)^S
\sim e^{iS\omega t_E}$. In the systems with time-reversal symmetry, the leading 
corrections come from the two loop diagram, $L=2,S=2$ and $V=1$, hence 
the leading 
Ehrenfest oscillation in $R_p(\omega)$ is $\sim e^{i2\omega t_E}$. In 
systems with broken time-reversal symmetry, the leading corrections come from three loop diagrams. The diagram 
with 
the Hikami box of $6$-point vertex ($L=3,S=3$ and $V=1$) leads to the 
Ehrenfest oscillation $\sim e^{i3\omega t_E}$. The diagram with two 
Hikami boxes ($L=3,S=4$ and $V=2$) leads to the Ehrenfest oscillation 
$\sim 
e^{i4\omega t_E}$.

\section{The Nonperturbative Part Of $R(\omega)$ In Systems With Broken Time-Reversal Symmetry}
\label{sec5}

In Sec.~\ref{sec3} and \ref{sec4}, we study the perturbative expansion around the saddle 
point $\Lambda$. As a result, it gives the weak localization corrections to the leading 
perturbative term of the universal Wigner-Dyson statistics. However, it is known that 
such an expansion cannot reproduce the Wigner-Dyson oscillation \cite{AA95}. Instead, Andreev and Altshuler found that such a nonperturbative 
term is controlled by other nontrivial saddle points \cite{AA95}. In this section, we apply their 
method to study the leading order corrections to the universal Wigner-Dyson oscillation in 
the region $\Delta\ll\omega\sim t_E^{-1}\ll t_{erg}^{-1}$.

\subsection{The global transformation}
\label{GT}

For 
GOE and GUE, the other saddle point is 
$-k\Lambda$ \cite{AA95}. To establish a perturbation theory, a 
trick, so-called global 
transformation on ${\bf H}$ is introduced, which maps the saddle 
point $-k\Lambda$ to $\Lambda$. That 
is, we perform a global coordinate transformation:
\begin{equation}
Q\rightarrow Q'=U_0^{-1}QU_0,\qquad T\rightarrow T'=TU_0,
\label{gloTra}
\end{equation} 
where $U_0\in {\bf H}$ satisfies $ -U_0^{-1}k\Lambda U_0=\Lambda$. 
It is important that the Jabobian is unity and $U_0$ does not depend on 
the coordinates. Under the transformation 
Eq.~(\ref{gloTra}), 
\begin{equation}
Q'=T'^{-1}\Lambda T'.
\label{degkLam}
\end{equation}
Thus, we can rewrite 
$R(\omega)$, 
Eq.~(\ref{R}) as
\begin{equation}
R(\omega)=\frac{1}{64}{\rm Re}\int DQ'\left(
\int dx_\parallel {\rm STr}\left[\Lambda Q'(x_\parallel)\right]
\right)^2 e^{-\tilde S_{eff}},
\label{Def_RkLam}
\end{equation}
where $\tilde S_{eff}$ is the action:
\ifpreprintsty \else\end{multicols} 
                  	\vspace*{-3.5ex} \hspace*{0.491\hsize} 
			\begin{minipage}{0.48\hsize}$\,$\hrule
			 \end{minipage}
\fi
\begin{equation}
\tilde S_{eff}=
\frac{\pi\nu}{2}\int dx_\parallel {\rm STr}\left[
-\frac{i\omega^+}{2}k\Lambda Q'-T'^{-1}\Lambda {\hat {\cal 
L}}T'+
\frac{1}{\tau_q}\left(
\frac{\partial Q'}{\partial \phi}
\right)^2
\right].
\label{newAction}
\end{equation}
\ifpreprintsty \else 
                  	\vspace*{-3.5ex} 
			\begin{minipage}{0.48\hsize}$\,$\hrule
			\end{minipage}\vspace*{1.5ex}
                  \begin{multicols}{2}
                   \fi

Although the findings in Ref.~\onlinecite{AA95} are asymptotic,  it 
is important that the 
perturbation 
near the nontrivial saddle points take into account nonzero mode 
approximations, which, in principle, cannot be done  by 
zero-dimensional 
supermatrix $\sigma$ model.  Indeed, for BNS$\sigma$M, the Gaussian 
approximation is not applicable \cite{KE02} in the nonuniversal region: 
$\omega\gg t_{erg}^{-1}$. However, we believe that the perturbation 
near saddle points is applicable in the universal region $\omega\ll 
t_{erg}^{-1}$, because 
the repetition problem \cite{BK96} is not essential in this case. In the ballistic 
case, the new region: $t_E^{-1}\sim\omega\ll 
t_{erg}^{-1}$ ,  appears, within which the relaxation of the momentum 
is not complete. As a result, one can not put $Q$ to be a constant 
matrix. 
The parameterization, Eqs.~(\ref{parameterization})-(\ref{condition}) 
still holds for $T'$. Furthermore, 
we can expand $B$ in terms of the Pauli's matrices:
\begin{equation}
B=\sum_k B_k\otimes \tau_{d,k}, \qquad {\bar B}=\sum_k {\bar 
B}_k\otimes \tau_{d,k}.
\label{B}
\end{equation}
The modes $k=0,3$ stand for the diffuson and the modes $k=1,2$ stand for the cooperon. 
They satisfy the same equation Eq.~(\ref{diffuson}) in the 
absence of magnetic fields. In this way, we call both of them diffuson 
modes in this paper unless special explanation. The modes $k=1,2$ are suppressed in systems with broken time-reversal symmetry 
due to the destruction of the interference between a trajectory and its time-reversed partner. In this paper, we will study only this case. 
Due to Eq.~(\ref{Tcon}), the matrices $B_k$ and ${\bar B}_k$ 
satisfy:
\begin{equation}
{\bar B}_k=kB_k^\dag,\qquad k=0,1,2,3.
\label{BBbar}
\end{equation}
$B_k$ may be parameterized as
\begin{equation}
B_k=\left(
\begin{array}{cc}
a_k & i\sigma_k\\
\eta_k^\ast & ib_k
\end{array}
\right),
\label{Bpar}
\end{equation}
where $a_k(1)$ and $b_k(1)$ are ordinary variables, while $\sigma_k(1)$ 
and $\eta_k(1)$ are Grassmann variables. Then we substitute 
Eqs.~(\ref{B})-(\ref{Bpar}) into Eq.~(\ref{Def_RkLam}) and keep the 
quadratic 
terms in $a_k, b_k, \sigma_k$ and $\eta^\ast_k$ to rewrite 
Eq.~(\ref{newAction}) as
\begin{equation}
\tilde S_{eff}^0=-2i\pi\omega+\tilde S^0.
\label{Seff0}
\end{equation}
\ifpreprintsty \else\end{multicols} 
                  	\vspace*{-3.5ex} \hspace*{0.491\hsize} 
			\begin{minipage}{0.48\hsize}$\,$\hrule
			 \end{minipage}
\fi
\begin{equation}
\tilde S^0=2\pi\int dx_\parallel\sum_{k=0,3} \left[
a_k^\ast(x_\parallel)\left(i\omega^++{\hat {\cal L}}_R\right)a_k 
(x_\parallel) +
b_k^\ast(x_\parallel)\left(-i\omega^++{\hat {\cal L}}_R\right) b_k 
(x_\parallel) +\sigma_k^\ast(x_\parallel) {\hat {\cal L}}_R\sigma_k(x_\parallel)+
\eta_k(x_\parallel) {\hat {\cal L}}_R\eta^\ast_k(x_\parallel)
\right].
\label{newS0}
\end{equation}
\ifpreprintsty \else 
                  	\vspace*{-3.5ex} 
			\begin{minipage}{0.48\hsize}$\,$\hrule
			\end{minipage}\vspace*{1.5ex}
                  \begin{multicols}{2}
                   \fi\noindent 
It is important to see that the new Gaussian action, $\tilde S^0$ does not include zero mode contribution of 
Grassmann fields. Thus, for a generic Gaussian integral with the action ${\tilde S}^0$  not to vanish, the 
prefactor must consist of all the independent zero mode Grassmann 
variables. In this way, by expanding $Q'$ in the prefactor up to the second order, 
and taking into account the zero mode bosonic and Grassmann fields only, 
the universal Wigner-Dyson oscillation is reproduced in Refs.~\onlinecite{ASAA96,AA95}.

\subsection{Quantum corrections}
\label{QC_ROSC}

The leading order corrections to the universal Wigner-Dyson oscillation has the amplitude 
$\sim (\Delta/\omega)^3$ at $\omega\gg \Delta$. To see this, one need to expand $T'^{-1}$ and 
keep $Q$ 
up to $P^4$ and take into account the first order effective interaction 
similarly to Eq.~(\ref{intOrt}). As a result, we obtain:

\begin{figure}
\begin{center}
\leavevmode
\epsfxsize=9cm
\epsfbox{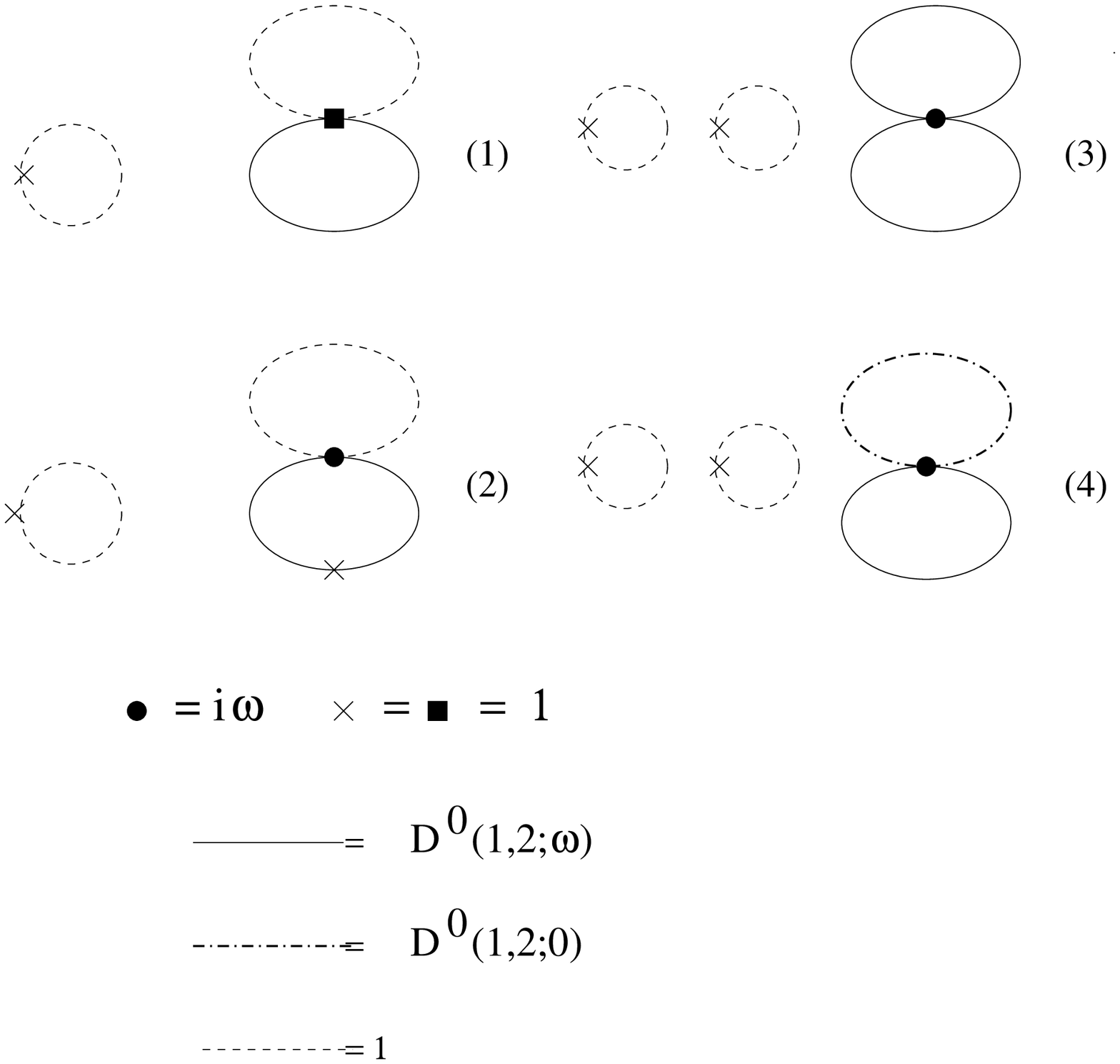}
\end{center}
\caption {Nonvanishing diagrams contribute to $R_{np}(\omega)$ in the 
order of 
$(\Delta/\omega)^3$ for systems with broken time-reversal symmetry. The cross and the square, 
$\sim {\rm Str} \left[P^2\right]$ and $\sim {\rm Str}\left[P^4\right]$, respectively, come from the expansion in the prefactor, see Eq.~(\ref{pre_expan2}). 
The dot, $\sim i\omega{\rm Str}\left[kP^4\right]$, comes from the effective interaction.
The solid line stands for the bosonic propagator, the dashed line stands for the zero mode Grassmann propagator, 
and the dash-doted line stands for the nonzero mode Grassmann propagator. In the figure, we put $\Delta =1$.
}
\label{Fig.3}
\end{figure}

\ifpreprintsty \else\end{multicols} 
                  	\vspace*{-3.5ex} \hspace*{0.491\hsize} 
			\begin{minipage}{0.48\hsize}$\,$\hrule
			 \end{minipage}
\fi
\begin{eqnarray}
\Delta R^{\rm u}_{np}(\omega)&=&\frac{1}{64}{\rm Re}\int DP e^{\frac{2i\pi\omega}{\Delta}}e^{-\tilde{S}^0}
\left[
-8\int d1 {\rm STr}\left[P^2(1)\right]\int d2 {\rm STr}\left[P^4(2)\right]
-4\int d1 {\rm 
STr}\left[P^2(1)\right]\int d2 {\rm STr}\left[P^2(2)\right]
\tilde{S}_{eff}^{int.}
\right],\nonumber\\
\tilde{S}_{eff}^{int.}&=&-\frac{\pi\nu}{2}i\omega\int dx_{\parallel} 
{\rm 
STr}
\left[kP^4(x_\parallel)\right].
\label{DelRuni}
\end{eqnarray}

To calculate Eq.~(\ref{DelRuni}), we need to separate the zero mode Grassmann fields from all 
the other fields and integrate them out first. Consequently we are left with a Gaussian 
integral involving the bosonic and nonzero mode Grassmann fields. Then we can factorize 
the integral into the product of the following pairs:
\begin{eqnarray}
4\pi\nu Z^{-1}\int D[a^\ast ab^\ast b\sigma^\ast 
\sigma\eta\eta^\ast]
a^\ast_k(1) a_k(2)e^{-\tilde{S}^0}={\cal D}^0(1,2;-\omega),
\label{newcont1}\\
4\pi\nu Z^{-1}\int D[a^\ast ab^\ast b\sigma^\ast 
\sigma\eta\eta^\ast]
b^\ast_k(1) b_k(2)e^{-\tilde{S}^0}={\cal D}^0(1,2;\omega),
\label{newcont2}\\
4\pi\nu Z^{-1}\int D[a^\ast ab^\ast b\sigma^\ast 
\sigma\eta\eta^\ast]
\sigma^\ast_k(1) \sigma_k(2)e^{-\tilde{S}^0}={\cal D}^0(1,2;0),
\label{newcont3}\\
4\pi\nu Z^{-1}\int D[a^\ast ab^\ast b\sigma^\ast 
\sigma\eta\eta^\ast]
\eta_k(1)\eta^\ast_k(2)e^{-\tilde{S}^0}={\cal D}^0(1,2;0),
\label{newcont4}
\end{eqnarray}   
\ifpreprintsty \else 
                  	\vspace*{-3.5ex} 
			\begin{minipage}{0.48\hsize}$\,$\hrule
			\end{minipage}\vspace*{1.5ex}
                  \begin{multicols}{2}
                   \fi\noindent 
where $k=0,3$, and ${\cal D}^0(1,2;0)$ satisfies the following equation:
\begin{equation}
{\hat {\cal L}}_R{\cal D}^0(1,2;0)=
2\pi\left[\delta(1-2)-1\right].
\label{D0}
\end{equation}
In the square integrable space, the overall factor, $Z$ can be approximated as  
\begin{equation}
Z=\int D[a^\ast ab^\ast b\sigma^\ast 
\sigma\eta\eta^\ast]e^{-\tilde{S}^0}
\simeq \frac{\Delta^2}{(4\pi)^2}\frac{1}{\omega^2}.
\label{Zu}
\end{equation}
We note that $Z\neq 1$ because $\tilde S^0$ is not supersymmtric and we do not integrate out zero mode Grassmann fields. 
Actually with nonzero modes integrated out, we come up with a factor $\left(1+a\right)$, 
where $a=\omega^2/\sum_k \lambda_k^2$ and the sum is over all the nonzero eigenvalues $\lambda_k$ of 
the Perron-Frobenius operator \cite{KM94,ASAA96,BMM01}. In the universal region $\omega\ll\lambda$, this 
factor crossovers to the universal value $1$. Furthermore, after integrating out 
zero mode bosonic fields, we obtain Eq.~(\ref{Zu}). The measure $D[a^\ast ab^\ast b\sigma^\ast 
\sigma\eta\eta^\ast]$ is on all the independent bosonic and nonzero mode 
Grassmann fields. Besides, one needs to take into account 
Eq.~(\ref{Tcon}) to reduce the number of integration variables by half.

Eq.~(\ref{DelRuni}) involves the Gaussian integral of bosonic and nonzero mode Grassmann fields, which can reduced 
into the sum with each term the products of pairs [Eq.~(\ref{newcont1}) to (\ref{newcont4})]. 
Eq.~(\ref{DelRuni}) corresponds to 4 nonvanishing diagrams shown in Fig.~\ref{Fig.3}. 
One should not confuse these diagrams with Fig.~\ref{fig1}. Here the 
Hikami box 
in the $8$-shaped diagram stands for the coupling between two 
diffusons, 
not a diffuson and a cooperon, as in the orthogonal case. It is 
important 
that such a diagram is exactly zero in the unitary case without the 
global transformation. In this way, the existence of these diagrams 
results 
from the supersymmetry breaking.

The first term in the r.h.s. of Eq.~(\ref{DelRuni}) gives 
\ifpreprintsty \else\end{multicols} 
                  	\vspace*{-3.5ex} \hspace*{0.491\hsize} 
			\begin{minipage}{0.48\hsize}$\,$\hrule
			 \end{minipage}
\fi
\begin{equation}
\Delta R^{\rm u}_{np,1}(\omega)=-2\frac{\Delta^3}{\pi^3\omega^2}{\rm 
Re}\left(
e^{2\pi i\omega/\Delta}\int d3
\left[\langle{\cal D}^0(3,3;-\omega)\rangle-
\langle{\cal D}^0(3,3;\omega)\rangle\right]
\right)
\simeq 
-4\frac{\Delta^3}{\pi^3\omega^3}{\rm 
Im} e^{2\pi i\omega/\Delta}
\label{REOuni3}
\end{equation}
which is the contraction shown in Fig.~\ref{Fig.3}(1). Above we use the fact that at $\omega\ll\lambda$, 
we need to take into account the zero mode only in the diffusons.

The second term in Eq.~(\ref{DelRuni}) has two nonvanishing contraction. One is shown in 
Fig.~\ref{Fig.3}(2), while the other is shown in Fig.~\ref{Fig.3}(3) and (4). In 
Fig.~\ref{Fig.3}(2), only the bosonic fields are involved. By using Eqs.~(\ref{newcont1}) 
and (\ref{newcont2}), we find it to be:
\begin{equation}
\Delta R^{\rm u}_{np,2}(\omega)=-\frac{\Delta^3}{\pi^3\omega^2}{\rm 
Re}\left(
(-i\omega)e^{2\pi i\omega/\Delta}\int d2\int d3
\left[\langle{\cal D}^0(3,2;-\omega){\cal D}^0(2,3;-\omega)\rangle+
\langle{\cal D}^0(3,2;\omega){\cal D}^0(2,3;\omega)\rangle\right]
\right).
\label{REOuni2}
\end{equation}
It should be emphasized that the order of the arguments appearing in the product of two diffusons above is essentially different from what 
we calculated in Sec.~\ref{two_diffus} [see Eq.~(\ref{Msolution})]. In fact, we can use Eq.~(\ref{Drelation}) to reduce the 
two diffusons in Eq.~(\ref{REOuni2}) into one diffuson. Thus, we can approximate the product of two diffusons as 
$1/(\pm i\omega)^2$ in the region $\omega\ll \lambda$.

Fig.~\ref{Fig.3}(3) differs from Fig.~\ref{Fig.3}(4) in that there the coupling occurs 
between bosonic fields. While Fig.~\ref{Fig.3}(4), the coupling occurs between the 
bosonic and nonzero mode Grassmann fields. By using Eqs.~(\ref{newcont1}) 
and (\ref{newcont2}), we find that Fig.~\ref{Fig.3}(3) corresponds to 
\begin{equation}
\Delta R^{{\rm u}}_{np,3}(\omega)=
-\frac{\Delta^3}{\pi^3\omega^2}{\rm Re}\left(
(-i\omega)e^{2\pi i\omega/\Delta}\int d3
\left[\langle{\cal D}^0(3,3;-\omega){\cal D}^0(3,3;-\omega)\rangle+
\langle{\cal D}^0(3,3;\omega){\cal D}^0(3,3;\omega)\rangle\right]
\right).
\label{REOuni1}
\end{equation}
By using Eqs.~(\ref{newcont1})-(\ref{newcont4}), we find that Fig.~\ref{Fig.3}(4) corresponds to 
\begin{equation}
\Delta R^{{\rm u}}_{np,4}(\omega)=
\frac{\Delta^3}{\pi^3\omega^2}{\rm Re}\left(
(-i\omega)e^{2\pi i\omega/\Delta}\int d3
\left[\langle{\cal D}^0(3,3;-\omega){\cal D}^0(3,3;0)\rangle+
\langle{\cal D}^0(3,3;\omega){\cal D}^0(3,3;0)\rangle\right]
\right).
\label{REOuni4}
\end{equation}
\ifpreprintsty \else 
                  	\vspace*{-3.5ex} 
			\begin{minipage}{0.48\hsize}$\,$\hrule
			\end{minipage}\vspace*{1.5ex}
                  \begin{multicols}{2}
                   \fi\noindent

In the disordered case, one is easy to check that $\Delta R^{\rm 
u}_{np}(\omega)$ 
vanishes using the zero mode diffuson: $\langle{\cal 
D}^0(1,2;\omega)\rangle=1/-i\omega^+$ and $\langle{\cal 
D}^0(1,2;0)\rangle=0$.  
However, it is not so in the crossover region $\Delta\ll\omega\sim t_E^{-1}\ll t_{erg}^{-1}$. 
Instead, as we see below, Eqs.~(\ref{REOuni3})-(\ref{REOuni4}) lead to the Ehrenfest oscillation with the 
period $(\Delta^{-1}+\alpha t_E)^{-1}$ with $\alpha$ a universal numerical factor.

\subsection{The Ehrenfest oscillation in $R_{np}^{\rm u}(\omega)$}
\label{Ehr_GUE_osc}

Now we show that in the crossover region, Eqs.~(\ref{REOuni3})-(\ref{REOuni4}) lead to the Ehrenfest oscillation, but with different dependence of the period on $t_E$ compared to the perturbative part.

First, according to Eq.~(\ref{MMsolution}), we can rewrite Eq.~(\ref{REOuni1}) as 
\begin{equation}
\Delta R^{{\rm u}}_{np,3}(\omega)=
\frac{\Delta^3}{\pi^3\omega^2}{\rm Re}\left[
e^{2\pi i\omega/\Delta}
\frac{\Gamma_2^2(\omega)+\Gamma_2^2(-\omega)}{-i\omega}
\right].
\label{REOuni1osc}
\end{equation}
In the case of $\lambda_2 \omega^2/\lambda^3\ll 1$, Eq.~(\ref{REOuni1osc}) is simplified as 
\begin{equation}
\Delta R^{{\rm u}}_{np,3}(\omega)=
-2\frac{\Delta^3}{\pi^3\omega^2}
\sin \frac{2\pi \omega}{\Delta}
\cos 2\omega t_E.
\label{REOuni1osc11}
\end{equation}

To calculate Eq.~(\ref{REOuni4}), first we understand ${\cal D}^0(3,3;0)$ in the way of ${\cal D}^0(3,3;0)=\lim_{\omega_1\rightarrow 0}\left[{\cal D}^0(3,3;\omega_1)+{\cal D}^0(3,3;-\omega_1)\right]/2$, where ${\cal D}^0(1,2;\omega_1)$ should be considered as the solution of the equation below:
\begin{equation}
\left(-i\omega^+_1+{\hat {\cal L}}_R\right){\cal D}^0(1,2;\omega_1)=
2\pi\left[\delta(1-2)-1\right].
\label{D0'}
\end{equation}
Then following the procedure of deriving Eq.~(\ref{Msolution}), we find that 
$\langle{\cal D}^0(3,3;\omega){\cal D}^0(3,3;\omega_1)\rangle=-\Gamma_2^2\left(\frac{\omega+\omega_1}{2}\right)/\omega^+\omega^+_1$. 
Thus we obtain:
\begin{equation}
\langle {\cal D}^0(3,3;\omega){\cal D}^0(3,3;0)\rangle
=-\frac{1}{\omega}\frac{\partial}{\partial\omega}\Gamma_2^2\left(\frac{\omega}{2}\right).
\label{OmegaO2}
\end{equation}
Substituting Eq.~(\ref{OmegaO2}) into Eq.~(\ref{REOuni4}), we find that $\Delta R^{{\rm u}}_{np,4}(\omega)$ is 
\begin{equation}
\Delta R^{\rm u}_{np,4}(\omega)=-2\frac{\Delta^3}{\pi^3\omega^3}
\sin \frac{2\pi \omega}{\Delta}
\omega {\rm Re}
\frac{\partial}{\partial \omega}\Gamma_2^2(\frac{\omega}{2}).
\label{REOuni44}
\end{equation}
In the case of $\lambda_2 \omega^2/\lambda^3\ll 1$, Eq.~(\ref{REOuni44}) is simplified as
\begin{equation}
\Delta R^{{\rm u}}_{np,4}(\omega)=
2\frac{\Delta^3}{\pi^3\omega^3}
\sin \frac{2\pi \omega}{\Delta}\omega t_E\sin \omega t_E.
\label{REOuni4411}
\end{equation}

Collecting Eqs.~(\ref{REOuni3}), (\ref{REOuni2}), (\ref{REOuni1osc}) and (\ref{REOuni44}) together, we find that the leading quantum corrections in the nonperturbative part of $R(\omega)$ is Eq.~(\ref{Intro_2_np}). 
Thereby $R_{np}^{\rm u}(\omega)$ acquires an oscillation 
correction with the period $(\Delta^{-1}+\alpha t_E)^{-1}$ with $\alpha$ 
a universal numerical factor. Actually we can find the similar results for systems with time-reversal symmetry, 
but the amplitudes are much smaller, which are proportional to 
$(\Delta/\omega)^5\cos (2\pi\omega/\Delta)$ or $(\Delta/\omega)^5\sin (2\pi\omega/\Delta)$.

In the low $\omega$ limit: $\omega t_E\ll 1$, $\Delta R_{np}^{\rm u}(\omega)\sim \Delta^3 t_E^2/\omega$. 
It is a small quantum correction to the universal Wigner-Dyson statistics. However, 
we are not able to find this term by using the method developed in Ref.~\onlinecite{KM94}. 
In the large $\omega$ limit: $\omega t_E\gg 1$, we see that the universal Wigner-Dyson statistics acquires 
a small universal correction $\sim (\Delta/\omega)^3 e^{2\pi i \omega/\Delta}$. But 
it is smaller than the nonuniversal correction, which is 
$\sim (\Delta/\lambda)^2 e^{2\pi i \omega/\Delta}$.

Taking into account Eqs.~(\ref{RpF}) and (\ref{Intro_2_np}), we can formally express 
$R(\omega)$ as the following :
\ifpreprintsty \else\end{multicols} 
                  	\vspace*{-3.5ex} \hspace*{0.491\hsize} 
			\begin{minipage}{0.48\hsize}$\,$\hrule
			 \end{minipage}
\fi
\begin{equation}
R(\omega)=1+a+\sum_n\left(\frac{\Delta}{\omega}\right)^n
C_n(\omega t_E)+1
\left[
\cos \frac{2\pi\omega}{\Delta}
\sum_n\left(\frac{\Delta}{\omega}\right)^n
D_n(\omega t_E)+
\sin \frac{2\pi\omega}{\Delta}
\sum_n\left(\frac{\Delta}{\omega}\right)^n
E_n(\omega t_E)
\right].
\label{Rfinal}
\end{equation} 
 	\ifpreprintsty \else 
                  	\vspace*{-3.5ex} 
			\begin{minipage}{0.48\hsize}$\,$\hrule
			\end{minipage}\vspace*{1.5ex}
                  \begin{multicols}{2}
                   \fi\noindent
The factor $a\sim \omega^2/\lambda^2$ arises from making the Gaussian approximation to the nonzero modes of the Perron-Frobenius operator \cite{ASAA96,BMM01}. 
We do not study this term in this paper because it does not lead to the Ehrenfest oscillations. For disordered systems, 
a similar term, $\sim \omega^2/E_{Th}^2$ was found in Ref.~\onlinecite{KM94}. 
In general, $C_n, D_n$ and $E_n$ are oscillating functions of $\omega t_E$ and 
proportional to $\Gamma_n(\omega)$. In the disordered limit, $C_n, D_n$ and $E_n$ become universal 
constants.

\section{Conclusions}
\label{sec7}

In this paper, we study the electron energy statistics in chaotic quantum dots 
with one macroscopic size $L$. For such systems, the inverse ergodic time and 
the Lyapunov exponent are of the same order: $t_{erg}^{-1}\sim\lambda\sim 
\frac{v_F}{L}$. Consequently, in the 
semiclassical limit, the Ehrenfest time $t_E$ opens an intermediate region 
$\Delta\ll t_E^{-1}\sim\omega\ll t_{erg}^{-1}$. 
We study the behavior of the two level correlation function, $R(\omega)$ in the universal region 
$\Delta\ll\omega 
\ll t_{erg}^{-1}$ in systems with broken time-reversal symmetry. 
Surprisingly, it is found that $R(\omega)$ deviates from the universal Wigner-Dyson 
statistics. Basically, we find that $R(\omega)$ acquires two types of 
oscillation corrections in 
the crossover region $t_E^{-1}\sim\omega\ll t_{erg}^{-1}$. The oscillation periods have different dependence 
on $t_E$. 
The Ehrenfest oscillations in the perturbative part have the period 
proportional to $t_E^{-1}$, while the Ehrenfest oscillations in the nonperturbative part have the period $(\Delta^{-1}+\alpha 
t_E)^{-1}$ with $\alpha$ being some universal numerical factor. These additional 
Ehrenfest oscillations are small corrections 
to the universal RMT results. In particular, for $R^{\rm 
u}_p(\omega)$ described by Eq.~(\ref{zeroRu}), the exact truncation at the term 
$\sim\omega^{-2}$ 
does not imply the disappearance 
of quantum corrections. Instead, it is due to the cancellations of 
contributions arising from Hikami boxes associating with different kinds of vertices at the limit $\omega t_E\rightarrow 0$. 
The Ehrenfest oscillations are just the reminiscence of such 
cancellations.

In this paper, only the leading Ehrenfest oscillation 
corrections are calculated. Actually there is a general expression, 
Eq.~(\ref{Rfinal}) in the region $\Delta\ll \omega\ll t_{erg}^{-1}$. The functions 
$\tilde C_n(x)$ [hence $C_n(x)$], $D_n(x)$ and $E_n(x)$ are oscillating at $\omega 
t_E\gtrsim 1$. 
In the limit $n\omega t_E\ll 1$, $\tilde C_n$ (hence $C_n$), $D_n$ and 
$E_n$ crossover to 
their universal values. In particular, 
in the GUE case, $C_2=-D_2=-1/2$, and all other coefficients vanish. 
The 
universal values of $\tilde C_n(x)$ [hence $C_n(x)$], $D_n(x)$ and $E_n(x)$ 
do not depend on the magnitude of the regularizer [the 
last term 
in Eq.~(\ref{SUSY action})]. But it depends on the order of how we take the 
limits: $\omega\rightarrow 0$ and $\tau_q\rightarrow\infty$. 
Actually we take the limit $\omega\rightarrow 0$ before 
$\tau_q\rightarrow\infty$. In contrast, the boundary of the region where the 
universality exists ($\omega\ll t_E^{-1}$) does logarithmically depend on the 
magnitude of $\tau_q$. The regularizer describe the coupling between 
classical 
trajectories. In this sense, it plays a role similar to that played 
by the interaction between particles in nonideal 
Bose gases. There the scaling theory describes the universal behavior 
at the 
transition point and the critical indices do not depend on 
interactions. However, 
the region for the scaling to be applicable does depend on the 
interaction. While in 
ideal 
gases, the scaling is absent. It is important that the results presented here 
hold only at $\Delta\ll\omega\ll\lambda$, 
where the saddle point approximation to Eqs.~(\ref{R}) and (\ref{SUSY 
action}) 
is possible.

We note that $\Gamma_n(\omega)$ is not analytical at 
$1/\tau_q$ 
in the large $\omega$ limit $t_E^{-1}\ll\omega<\lambda$. Actually 
following from 
Eqs.~(\ref{Gamma2}) and (\ref{Gamma3def}), we find that 
\begin{equation}
\Gamma_n(\omega)
\sim\left(\frac{1}{\tau_q}\right)^{\omega^2/(\lambda^3/\lambda_2)+i\omega/\lambda}, 
\qquad t_E^{-1}\ll\omega<\lambda.
\label{Gamman}
\end{equation}
From this, we see that corrections to $R(\omega)$, predicted by the Gutzwiller formula, 
are proportional to $\tau_q^{-\alpha}, 0<\alpha\ll 1$. In this way, any attempts of establishing a $1/\tau_q$ expansion 
are 
prohibited.

At higher $\omega$, $\omega\gtrsim\lambda\sim 
t_{erg}^{-1}$, the nonzero mode contributions turn out to be 
important. 
In this case, the saddle point approximation is no longer applicable. A 
refined technique 
is desired to study the oscillation, claimed in 
Ref.~\onlinecite{BMM01} 
for generic chaotic quantum dots.

In this paper, we consider chaotic 
quantum dots where $\lambda\sim t_{erg}^{-1}\sim 
\frac{v_F}{L}$. Actually, it is possible that $\lambda\gg t_{erg}^{-1}$. 
In the latter case, the dot has large enough size to contain a lot of 
classical impurities inside. In this case, the Lyapunov exponent $\lambda\sim nav_F$, where 
$n$ is the concentration of impurities and $a$ is the size of classical impurities \cite{Dor95}. 
The ergodic time (Thouless time) $t_{erg}=L^2/D\gg\lambda^{-1}$. 
Here $D$ is the classical diffusion constant. In this case, contributions from nonzero modes 
are possible 
to be described by the saddle point approximation. The similar 
Ehrenfest 
oscillations are expected to exist at $t_{erg}^{-1}\lesssim\omega\sim 
t_E^{-1}<\lambda$. 
We leave this work in the future.

Based on the present work, it is unclear whether the BGS conjecture \cite 
{BGS84} may work in the universal region $\omega\ll t_{erg}^{-1}$ for generic quantum 
chaotic systems. In other words, are the fluctuations of energy levels described by the universal 
Wigner-Dyson statistics in such systems? According to Eq.~(\ref{Rfinal}), we 
point out that in chaotic quantum billiards a new scale, the Ehrenfest time appears, 
which results in the new types of oscillations 
at $\omega\sim 
t_E^{-1}$. 
These oscillations are not expected by either the RMT, zero mode 
nonlinear supermatrix $\sigma$ model or the Gutzwiller trace formula. 
Secondly, despite of the appearance of the Ehrenfest 
oscillations 
in the region $\omega\sim t_E^{-1}\gg\Delta$, their amplitudes are small. 
In this way, $R(\omega)$ is still dominated by the universal Wigner-Dyson statistics. Finally, 
the behavior of $R(\omega)$ at $\omega\sim 
\Delta$ remains an open problem. Indeed, in the present work we use the saddle 
point 
approximation to take into account the zero mode contributions. Furthermore, the perturbation theory 
near the saddle points predicts the regions for the terms in the expansion Eq.~(\ref{Rfinal}) to approach their universal 
limits. However, we are not aware 
of 
the limit of the sum of such an asymptotic expansion.

\section*{Acknowledgements}

We thank I. L. Aleiner, K. B. Efetov, A. Kamenev, B. I. Shklovskii and M. G. Vavilov for a lot of fruitful discussions. The work is supported by NSF grant 
No. 0120702.

\appendix

\section{The derivation of Equation~(\ref{DDD})}
\label{Deriva_DDD}

In this Appendix we prove Eq.~(\ref{DDD}). We will follow the general 
method 
developed in Ref.~\onlinecite{AL96}. In the discussions below, we 
ignore the 
small regularizer in Eq.~(\ref{3Dmotion'}) for the moment. Moreover, we 
turn to the time representation. Then employing the change of variables 
\begin{eqnarray}
&&z=(z_1,z_2),\qquad \alpha=(\alpha_1,\alpha_2),\nonumber\\
&&z_1=\ln\left[\theta_1^2+\left(\frac{\rho_1}{L}\right)^2\right]^{1/2},\qquad
\alpha_1=\arctan \frac{\theta_1 L}{\rho_1},\nonumber\\
&&z_2=\ln\left[\theta_2^2+\left(\frac{\rho_2}{L}\right)^2\right]^{1/2},\qquad
\alpha_2=\arctan \frac{\theta_2 L}{\rho_2},
\label{zalpha'}
\end{eqnarray}
we rewrite Eq.~(\ref{3Dmotion'}) as (after averaging over the 
coordinates of the center of mass)
\begin{eqnarray}
&&\left(
\frac{\partial}{\partial t}+{\hat D}_1+{\hat D}_2
\right)N(z,\alpha;2,2)=0,\nonumber\\
&&{\hat D}_1=-B_1(t)\sin 2\alpha_1\frac{\partial}{\partial z_1}+
\left[
B_1(t)\cos 2\alpha_1+B_2(t)
\right]\frac{\partial}{\partial \alpha_1},\nonumber\\
&&{\hat D}_2=-B_1(t)\sin 2\alpha_2\frac{\partial}{\partial z_2}+
\left[
B_1(t)\cos 2\alpha_2+B_2(t)
\right]\frac{\partial}{\partial \alpha_2},\nonumber\\
&&B_{1,2}(t)=\frac{v_F}{2L}\mp\frac{L}{2p_F}\frac{\partial^2U}{\partial 
R_\bot^2}\arrowvert_{{\bf R}={\bf R}(t,{\bf R}_0)}.
\label{NNN}
\end{eqnarray}
Introducing the change of variables 
\begin{equation}
x=\frac{z_1+z_2}{2},\qquad y=z_1-z_2,
\label{xy}
\end{equation} 
we obtain: 
\begin{eqnarray}
&&{\hat D}_1+{\hat D}_2=\nonumber\\
&&-\frac{1}{2}B_1(t)\left(\sin 2\alpha_1+
\sin 
2\alpha_2\right)
\frac{\partial}{\partial x}+B_1(t)\left(\sin 2\alpha_1-\sin 
2\alpha_2\right)
\frac{\partial}{\partial y}\nonumber\\
&&+\left[
B_1(t)\cos 2\alpha_1+B_2(t)
\right]\frac{\partial}{\partial \alpha_1}+\left[
B_1(t)\cos 2\alpha_2+B_2(t)
\right]\frac{\partial}{\partial \alpha_2}.\nonumber\\
\label{D1+D2}
\end{eqnarray}

The formal solution of Eq.~(\ref{NNN}) is 
\begin{eqnarray}
&&N(t;x,y;2,2)=\nonumber\\
&&\exp \left[C_1(t,\alpha_{1,2})\frac{\partial}{\partial x}\right]N
(0;x,y,{\hat {\alpha}}^0(\alpha,t);2,2),
\label{NNNfs}
\end{eqnarray}
where $C_1$ is
\begin{equation}
C_1(t,\alpha_{1,2})=\frac{1}{2}\int_0^t dt_1 B_1(t_1)
\left(
\sin 2{\hat \alpha_1}
+\sin 2{\hat \alpha_2}
\right).
\label{C1C2}
\end{equation}
The functions of $y(t)$ and ${\hat 
\alpha}_{1,2}(\alpha^0_{1,2},t)$ 
satisfy the following equations:
\begin{equation}
\frac{\partial y}{\partial t}=
B_1(t)\left(\sin 2\alpha_1-\sin 2\alpha_2\right),
\label{xymotion}
\end{equation}
\begin{eqnarray}
&&\frac{\partial}{\partial t}{\hat \alpha}_{1,2}=
B_1(t)\cos 2{\hat \alpha}_{1,2}+B_2(t), \nonumber\\
&&{\hat \alpha}_{1,2}(\alpha_{1,2}^0,0)=\alpha_{1,2}^0,
\label{alphamot}
\end{eqnarray}
with ${\hat \alpha}_{1,2}$ implicitly defined by 
\begin{equation}
{\hat \alpha}_{1,2}\left[
{\hat \alpha}_{1,2}^0(\alpha_{1,2},t),t
\right]={\hat \alpha}_{1,2}.
\label{alphacon}
\end{equation}
The formal solution of $y(t)$ is 
\begin{equation}
y(t)=y(0)+\int_0^t dt_1 B_1(t_1)\left(\sin 2\alpha_1-\sin 2\alpha_2\right)
\label{ys'}
\end{equation}
with $y(0)$ being the initial condition.

We are interested in the region 
$t\gtrsim\lambda^{-1}$. 
In this case, ${\hat \alpha}_{1,2}\left({\hat \alpha}_{1,2}^0,t\right)$ 
become self-averaging over the coordinates of the center of mass and no 
longer depend on the initial conditions. Consequently, at such large 
times, $y(t)$ reaches a 
constant depending on the initial value [see Eq.~(\ref{ys'})]. In this way, $N(t;x,y;2,2)$ 
depends on $y$ parametrically and has no 
$\alpha_{1,2}$ dependence. The evolution of $N(t;x,y;2,2)$ is governed 
by the Fokker-Planck 
type equation:
\begin{equation}
\left[\frac{\partial}{\partial t}-{\cal F}\left(
\frac{\partial}{\partial x}
\right)
\right]N(t;x,y;2,2)=0,
\label{NtF}
\end{equation}
where ${\cal F}\left(\frac{\partial}{\partial x}\right)$ is defined as 
\begin{equation}
{\cal F}\left(\frac{\partial}{\partial 
x}\right)=\lim_{t\rightarrow\infty}\frac{1}{t}\ln 
\int \frac{d{\bf n}d{\bf R}}{2\pi A}\exp \left[
B(t)\left(
\frac{\partial}{\partial x}
\right)
\right].
\label{Fdef}
\end{equation}
Here we keep in mind the important fact that $\int_0^t dt B_1(t)\sin 
\left[
2{\hat \alpha_{1,2}}(\alpha^0_{1,2},t)\right]$ does not depend on 
$\alpha_{1,2}$ 
due to the self-averaging at large times, so that 
$\int_0^t dt B_1(t)\sin \left[2{\hat \alpha_1}(\alpha^0_1,t)\right]=
\int_0^t dt B_1(t)\sin \left[2{\hat \alpha_2}(\alpha^0_2,t)\right]$. In this 
way, 
${\cal F}$ has no $\frac{\partial}{\partial y}$ dependence. 
\begin{equation}
B(t)=\int_0^t dt B_1(t)\sin \left[
2{\hat \alpha}(\alpha^0,t)
\right],
\label{Baverage}
\end{equation}
where $\alpha$ can take the subscript either $1$ or $2$.

Because $N(t;x,y;2,2)$ slowly varies over $x$, we expand ${\cal F}$ up 
to the second order 
in $\frac{\partial}{\partial x}$ to get:
\begin{equation}
{\cal F}\left(
\frac{\partial}{\partial x}
\right)=
\lambda\frac{\partial}{\partial x}+
\frac{\lambda_2}{2}\frac{\partial^2}{\partial x^2},
\label{F2nd'}
\end{equation}
where 
\begin{equation}
\lambda =\lim_{t\rightarrow\infty}\frac{1}{t}
\int \frac{d{\bf n}d{\bf R}}{2\pi A}B(t)
\label{aaa}
\end{equation}
and 
\begin{equation}
\lambda_2 =\lim_{t\rightarrow\infty}\frac{1}{t}
\{
\left[\int \frac{d{\bf n}d{\bf R}}{2\pi A}B^2(t)\right]-
\lambda^2 t^2
\}.
\label{aaa'}
\end{equation}
It is important that here $\lambda$ and $\lambda_2$ are exactly the 
same as 
those appearing in $\Gamma_2(\omega)$ because $B(t)$ is the same as 
that in Ref.~\onlinecite{AL96}. 
This is not surprising because mathematically, $\lambda$ is determined 
by the eigenvalue 
of the stability matrix of a trajectory \cite{LL}. 
Returning to the frequency representation, we find 
Eq.~(\ref{DDD}).

\ifpreprintsty\newpage

\else\end{multicols}\fi
\begin{references}
\bibitem{AL97'}We notice that in Ref.~\onlinecite{AL97}, it is a 
mistake to put $I_1+I_2$ [see Eqs.~(\ref{dort1}) and (\ref{dRort2})] to be zero. Consequently, the 
result there is in a slight difference from what we show in Eq.~(\ref{Intro_1}) and can not be expressed as the second derivative of the free energy 
$F_2(\omega)$.
\bibitem{DC}In the language of quantum disorders, the diffuson stands for the interference between a path and 
itself, and the cooperon stands for the interference between a path and its time-reversed partner.
\end{references}
\end{document}